# Non-reciprocal Components Based on Switched Transmission Lines

Aravind Nagulu, Tolga Dinc, Zhicheng Xiao, Mykhailo Tymchenko, Dimitrios Sounas, Andrea Alù, and Harish Krishnaswamy

***Abstract*—Non-reciprocal components, such as isolators and circulators, are critical to wireless communication and radar applications. Traditionally, non-reciprocal components have been implemented using ferrite materials, which exhibit non-reciprocity under the influence of an external magnetic field. However, ferrite materials cannot be integrated into IC fabrication processes, and consequently are bulky and expensive. In the recent past, there has been strong interest in achieving non-reciprocity in a non-magnetic IC-compatible fashion using spatio-temporal modulation. In this paper, we present a general approach to non-reciprocity based on switched transmission lines. Switched transmission lines enable broadband, lossless and compact non-reciprocity, and a wide range of non-reciprocal functionalities, including non-reciprocal phase shifters, ultra-broadband gyrators and isolators, frequency-conversion isolators, and high-linearity/high-frequency/ultra-broadband circulators. We present a detailed theoretical analysis of the various non-idealities that impact insertion loss and provide design guidelines. The theory is validated by experimental results from discrete-component-based gyrators and isolators, and a 25 GHz circulator fabricated in 45 nm SOI CMOS technology.**

## I. Introduction

Non-reciprocal components, such as gyrators, isolators, and circulators, have numerous applications in the fields of wireless communication and radar. Frequency-modulated continuous-wave (FMCW) radars typically operate in simultaneous-transmit-and-receive (STAR) mode, and a circulator is critical to enable the transmitter and the receiver to share the same antenna and avoid saturation of the receiver. Full-duplex wireless is an emerging wireless communication paradigm which has drawn significant research interest in recent years [1]–[9] due to its potential to double the spectral efficiency in the physical layer and offer numerous other benefits in higher layers. Unlike traditionally-used time-division or frequency-division duplexing schemes, in full-duplex wireless, the transmitter and the receiver *operate at the same time and at the same frequency*. Once again, the circulator is critical in allowing the the transmitter and the receiver to share the same antenna. Isolators are commonly used to protect power amplifiers from reflections at the antenna interface. The gyrator was postulated by Tellegen as the fifth linear circuit element after the resistor, capacitor, inductor and transformer [10], providing a non-reciprocal phase equal to $\pi$, and is a basic non-reciprocal element which can be used to realize arbitrary passive non-reciprocal circuits [11], [12].

A.N. and H.K. are with the Department of Electrical Engineering, Columbia University, New York, NY 10027, USA. e-mail: hk2532@columbia.edu.
Z.X., M.T., D.S., A.A. are with the Department of Electrical and Computer Engineering, University of Texas at Austin, TX 78712, USA.
A.A. is also with the Advanced Science Research Center, City University of New York, New York, NY 10031, USA.

Lorentz Reciprocity states that any linear and time-invariant medium with symmetric permittivity and permeability tensors is reciprocal. Historically, reciprocity has been broken by biasing magnetic (ferrite) materials using permanent magnets [13], [14]. However, magnetic materials are incompatible with IC fabrication technology, and therefore magnetic non-reciprocal components tend to be bulky and expensive. IC-compatible circulators have been proposed using the inherent non-reciprocal behavior of active voltage-/current-biased transistors [15]–[18]. However, active non-reciprocal components are severely limited by the linearity and noise performance of the active transistors, and are therefore not suitable for wireless applications which demand stringent performance on those fronts [19]. Nonlinearity can be used to violate reciprocity, and has been extensively studied in the optical domain [20]–[23], but these techniques have limited utility in wireless communication applications due to their stringent linearity constraints.

Recently, exciting research efforts have been made to break reciprocity using spatio-temporal modulation [24]–[27], with permittivity being the material parameter that is modulated. In [24], a transmission line is periodically loaded with varactors, and a traveling-wave modulates the varactors along the line. The unidirectional modulation signal imparts a direction-dependent frequency conversion to the input signal as it travels along the line. Thus, forward and reverse traveling signals are separated in frequency, and can be isolated from each other using a frequency diplexer. In this structure, the length of the transmission line required is inversely proportional to the modulation contrast. In general, however, permittivity modulation is weak, with varactors exhibiting a typical $C_{max}/C_{min}$ ratio of 2-4 on chip. As a result, a large form-factor is required to achieve strong non-reciprocity. Furthermore, varactors exhibit very poor quality factor as the operation frequency is increased to millimeter-wave frequencies, resulting in higher insertion loss. In [25], [26], it was demonstrated that traveling-wave modulation in a resonant ring results in angular momentum biasing and strong non-reciprocity without frequency conversion. The ring can be miniaturized using lumped components, but this degrades the operating bandwidth. Most importantly, as before, the limitations of permittivity modulation using varactors, namely limited modulation contrast and the degradation of the quality factor as frequency increases, remain.

On the other hand, conductivity can be very efficiently and strongly modulated in semiconductor media, with CMOS transistors exhibiting ON/OFF condutance ratios as high as $10^3$ - $10^5$ [28]. It has been shown that by modulating the conductivity, i.e. switching transistors, on either side of a delay medium, extremely strong, low-loss, and broadband non-reciprocity can be achieved within a small form factor [29]–[34]. The first demonstration involved staggered (i.e., phase-shifted) switching of transistors across a capacitor bank (commonly called an N-path filter [35], [36]) [29]. The phase-shifted N-path filter realizes an electrically-infinitesimal gyrator, which was then embedded within a ring to realize a compact circulator. The placement within the ring was further optimized to suppress the voltage swing at the gyrator for transmitter-port excitations, enhancing the linearity and power handling. However, this approach suffers from low bandwidth because the capacitors do not provide a true time delay. In addition to this, N-path filters are not realizable at millimeter-wave frequencies due to their stringent clocking requirements, as they require multiple phases of a clock at the frequency of operation. When the delay medium is replaced by a transmission line of appropriate length, non-reciprocity can be observed over much wider bandwidths. In addition to this, the modulation frequency can be greatly reduced relative to the operating frequency [30], [31]. These concepts were leveraged to realize the first millimeter-wave (25 GHz) passive non-magnetic circulator in CMOS technology in [30], [31]. Similar concepts were used in [32] to demonstrate an ultra-broadband circulator operating from 200 KHz to 200 MHz using discrete switches and co-axial cable delays, and in [33] to implement a 0.2$\mu$m GaN HEMT MMIC circulator operating up to 1GHz.

This paper presents an exhaustive set of various non-reciprocal structures possible by using spatio-temporal conductivity-modulation across transmission line delays, including an arbitrary phase-non-reciprocal element, an ultra-broadband gyrator, a frequency-conversion isolator, an ultra-broadband isolator, and ultra-broadband/high-linearity/high-frequency circulators. A detailed analysis for estimating the transmission losses is presented, along with results from a 25 GHz 45 nm SOI CMOS circulator prototype as a case study. This analysis can aid in choosing between different implementation technologies, fabrication processes, and non-reciprocal element topologies, and in performance optimization during the design phase. The rest of the paper is organized as follows. Section II discusses various phase-non-reciprocal configurations. Section III discusses various isolator and circulator topologies. Section IV contains a detailed analysis for estimating losses. Section V considers case studies implemented using discrete switches and co-axial cables, while Section VI details the 25 GHz 45 nm SOI CMOS circulator case study. Section VII concludes the paper.

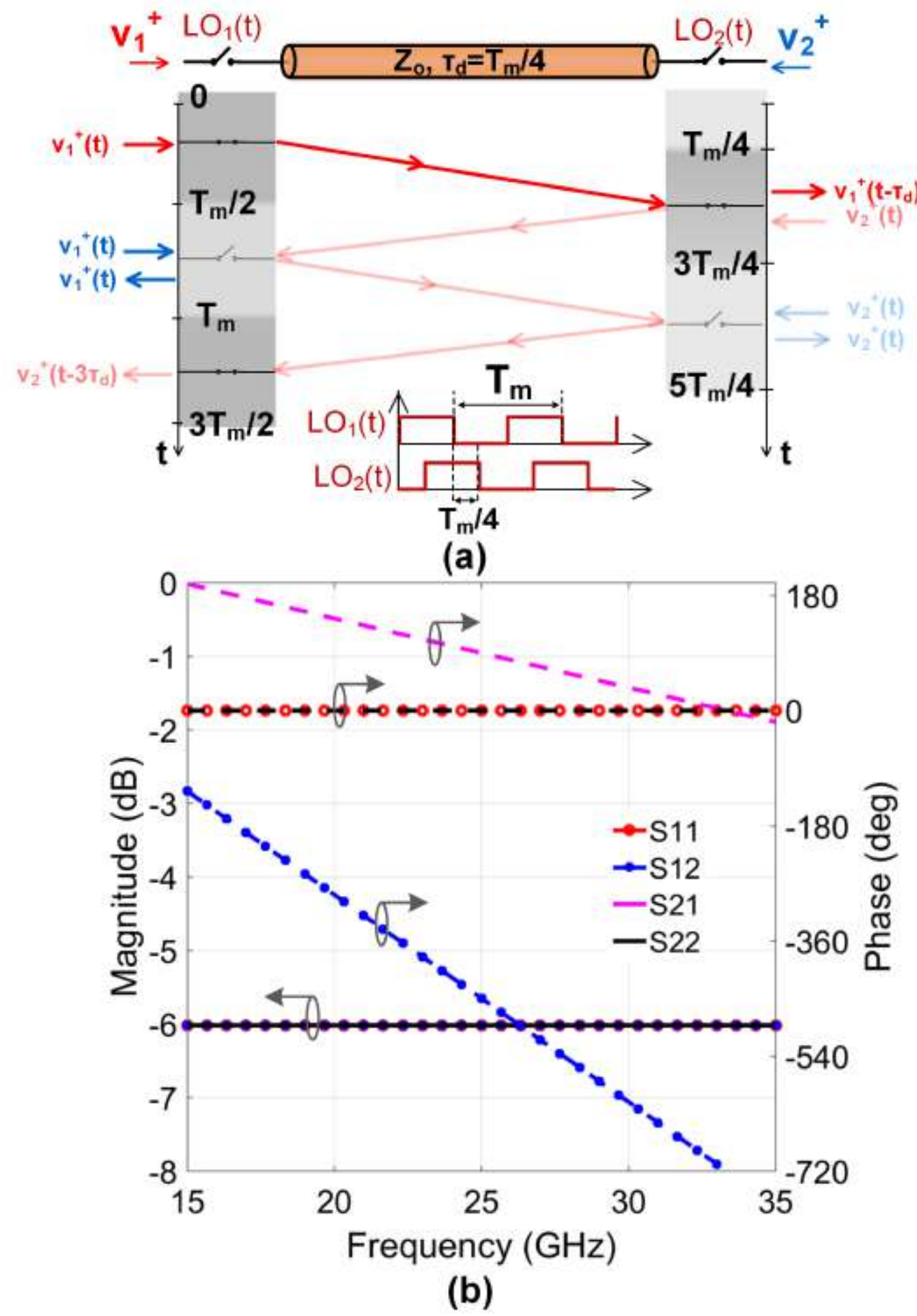

Fig. 1. Single-ended phase-non-reciprocal element: (a) circuit diagram and operation, and (b) fundamental-to-fundamental scattering parameters for $f_m$ = 8.33 GHz.

## II. Phase Non-Reciprocity

### A. Single-ended Phase-Non-Reciprocal Element

Fig. 1(a) depicts the circuit diagram and the operation of a single-ended phase-non-reciprocal element. It consists of a transmission-line segment whose characteristic impedance is equal to the port impedance and is sandwiched between two switches which are modulated using signals $LO_1(t)$ and $LO_2(t)$. The switches toggle between zero resistance (ON state) and infinite resistance (OFF state) when their modulation signals change between high and low values (1 and 0), respectively. The modulation signals $LO_1(t)$ and $LO_2(t)$ are $50\%$ duty cycle square-wave signals with a period of $T_m$. The modulation signal of the right-hand-side switch, $LO_2(t)$, is delayed with respect to that of the left-hand-side switch, $LO_1(t)$, by $T_m/4$, which is also the propagation delay of the transmission line.

The operation of this structure can be explained using time-domain analysis. For signals travelling from left to right, any arbitrary signal incident at port 1 when $LO_1(t)$ is high will be transmitted to port 2 in one pass, experiencing a delay of $T_m/4$ because the propagation delay of the transmission line is equal to the delay between the modulation signals. The signal incident at port 1 gets reflected when $LO_1(t)$ is low.

However, for signals travelling from right to left, the signal incident at port 2 when $LO_2(t)$ is high will be transmitted to port 1 after three passes, experiencing a delay of $3T_m/4$, because at the end of the first two passes, the signal sees an open termination due to an OFF state switch and experiences a total reflection. When $LO_2(t)$ is low, the signal incident at port 2 gets reflected. As a result of the additional reflections for incident signals from port 2 compared to port 1, the structure exhibits a different phase response for the signals travelling from left to right and right to left, making the structure non-reciprocal in phase. The behavior of the structure can be expressed in the time domain as

$$v_1^-(t) = v_1^+(t)\left(1 - LO_1(t)\right) + v_2^+\left(t - \tfrac{3T_m}{4}\right) LO_2\left(t - \tfrac{3T_m}{4}\right), \tag{1}$$

$$v_2^-(t) = v_1^+\left(t - \tfrac{T_m}{4}\right) LO_1\left(t - \tfrac{T_m}{4}\right) + v_2^+(t)\left(1 - LO_2(t)\right). \tag{2}$$

In general, LPTV circuits are represented using harmonic transfer functions (HTFs) which capture frequency conversion effects [37], [38]. For simplicity, here we only show the fundamental-to-fundamental S-parameter matrix by taking Fourier transform of the time-domain equations:

$$S = \begin{pmatrix} \frac{1}{2} & \frac{1}{2}e^{-j\omega\frac{3T_m}{4}} \\ \frac{1}{2}e^{-j\omega\frac{T_m}{4}} & \frac{1}{2} \end{pmatrix}. \tag{3}$$

One-fourth of the power incident into the port is transmitted to the other port, one-fourth of the power is reflected due to imperfect matching, and the remaining half is converted to other frequencies (i.e., intermodulation products) due to the switching action. Fig. 1(b) shows the fundamental-to-fundamental S parameters around 25 GHz for $f_m$=8.33 GHz.

### B. Balanced Configuration: Arbitrary Phase-Non-Reciprocal Element

Harmonic conversion and the effect of imperfect matching of the single-ended phase-non-reciprocal element can be easily overcome by adding another parallel branch switched with complementary clocks $\overline{LO_1}(t)$ and $\overline{LO_2}(t)$, as shown in Fig. 2(a). The incident signal at port 1 (port 2) travels through the first transmission line when $LO_1(t)$ ($LO_2(t)$) is high, and the second when $LO_1(t)$ ($LO_2(t)$) is low, making the structure completely matched with lossless transmission. The behavior of the structure can be expressed in the time domain as

$$v_1^-(t) = v_2^+\left(t - \frac{3T_m}{4}\right), \tag{4}$$

$$v_2^-(t) = v_1^+\left(t - \frac{T_m}{4}\right). \tag{5}$$

By taking Fourier transform of the time-domain equations, the S-parameter matrix can be calculated as

$$S = \begin{pmatrix} 0 & e^{-j\omega\frac{3T_m}{4}} \\ e^{-j\omega\frac{T_m}{4}} & 0 \end{pmatrix}. \tag{6}$$

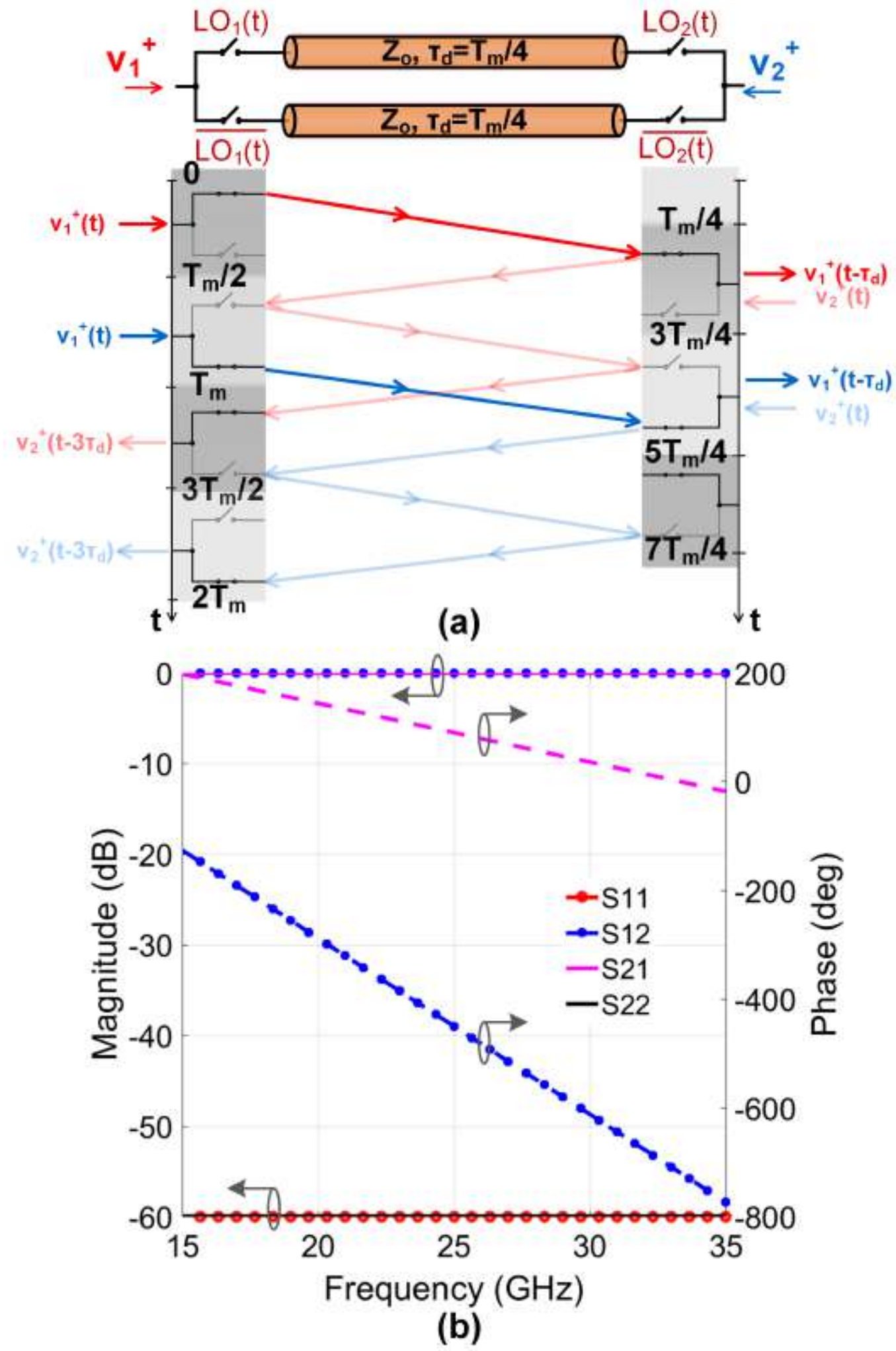

Fig. 2. Balanced arbitrary phase-non-reciprocal element: (a) circuit diagram and operation, and (b) scattering parameters for $f_m$ = 8.33 GHz.

Interestingly, in this balanced configuration, intermodulation products generated in the first path and the second path will have opposite phase and cancel each other. Therefore, while the structure is internally linear and periodically-time-varying (LPTV), at its ports, it features no frequency conversion and appears linear and time-invariant (LTI).

At the odd multiples of the modulation frequency, $\omega = (2n+1)\omega_m$, where $\omega_m = 2\pi/T_m$ and $n$=0,1,2,3..., the phases from left to right and right to left are $-(2n+1)\pi/2$ and $-3(2n+1)\pi/2$ respectively. In this case, the structure behaves as a gyrator, providing a non-reciprocal phase equal to $\pi$ [10]. Fig. 2(b) shows the S parameters around a gyration frequency of 25 GHz for $f_m$=8.33 GHz. In addition to this, at any given input frequency, a value of the modulation frequency $\omega_m$ can be chosen to realize a non-reciprocal phase difference of arbitrary value, $\Delta\Phi = \pi\left(\omega/\omega_m\right)$. Of course, reconfiguring the value of this non-reciprocal phase shift after implementation requires reconfiguring the transmission line so that its delay matches the modulation frequency change.

While ideally lossless, in practice, the transmission loss will be limited by the quality factor of the transmission line and non-zero switch resistance. The bandwidth of operation will be limited by the dispersion characteristics of the transmission line, switch parasitic capacitance, and rise/fall time of the modulation clocks. In the presence of a finite quality factor in the transmission lines, signals incident at port 2 experience more loss because they get transmitted after three passes when compared to signals incident at port 1 which get transmitted after one pass.

### C. Doubly-balanced Configuration: Ultra-broadband Gyrator

Even though the balanced structure provides non-reciprocal phase response over an infinite bandwidth, it behaves as a gyrator only at discrete frequencies. In addition to this, in a practical implementation, the transmission loss in forward and reverse directions will be imbalanced. Another interesting variant is a doubly-balanced non-reciprocal element, as shown in Fig. 3(a), which exploits differential signaling for both the modulation signal as well as the input signal. It consists of a differential transmission line, with a characteristic impedance equal to the differential port impedance, which is sandwiched between two doubly-balanced switch sets (often called Gilbert-quad mixers). A Gilbert-quad switch set consists of four switches where two switches connect the input and output directly when the modulation signal is high and the other two switches (crisscrossed) swap the signal polarity when the modulation signal is low. The switch sets on the left and right of the transmission line are driven by modulation signals $LO_1(t)$ and $LO_2(t)$ respectively. Similar to the balanced case, the modulation signal of the right-hand side switch, $LO_2(t)$, is delayed with respect to that of the left-hand side switch, $LO_1(t)$, by a value $T_m/4$, which is also the propagation delay of the transmission line.

In this configuration, signals travelling from left to right, incident at port 1, get transmitted to port 2 without any sign inversion in the first half cycle, and with two sign inversions that cancel each other in the second half cycle. On the other hand, signals travelling from right to left, incident at port 2, get transmitted to port 1 with a sign inversion from the left-hand-side mixer in the first half cycle and a sign inversion from the right-hand-side mixer in the second half cycle. The behavior of the structure can be expressed in the time domain as shown in (7) and (8), where $m(t)$ is the multiplication factor due to the switching action of a quad-mixer, which is a square wave signal between +1 and -1 with 50% duty cycle and period of $T_m$. By taking Fourier transform of the time-domain equations, the S-parameter matrix can be calculated as

$$S = \begin{pmatrix} 0 & -e^{-j\omega\frac{T_m}{4}} \\ +e^{-j\omega\frac{T_m}{4}} & 0 \end{pmatrix}. \quad (9)$$

As expected, this doubly-balanced structure also appears LTI externally while being internally LPTV. Most importantly, this structure describes an ideal gyrator over an infinite bandwidth. Fig. 3(b) shows the S parameters around 25 GHz for $f_m$=8.33 GHz. Ideally, zero insertion loss and a perfect non-reciprocal phase shift of $\pi$ are possible over infinite bandwidth at arbitrarily small sizes through appropriate increase of the modulation frequency. In practice, transmission losses will be limited by the quality factor of the transmission lines and non-zero switch resistance, the bandwidth of operation by the dispersion characteristics of the transmission line, switch parasitic capacitance, and rise/fall times of the modulation clock, and the size by the practicalities of switching solid-state switches at increasing modulation frequency and the associated power consumption. It should also be mentioned that in the presence of losses, this structure continues to behave as a perfect gyrator with symmetric losses, because the phase non-reciprocity is achieved due to sign inversion as opposed to path length differences.

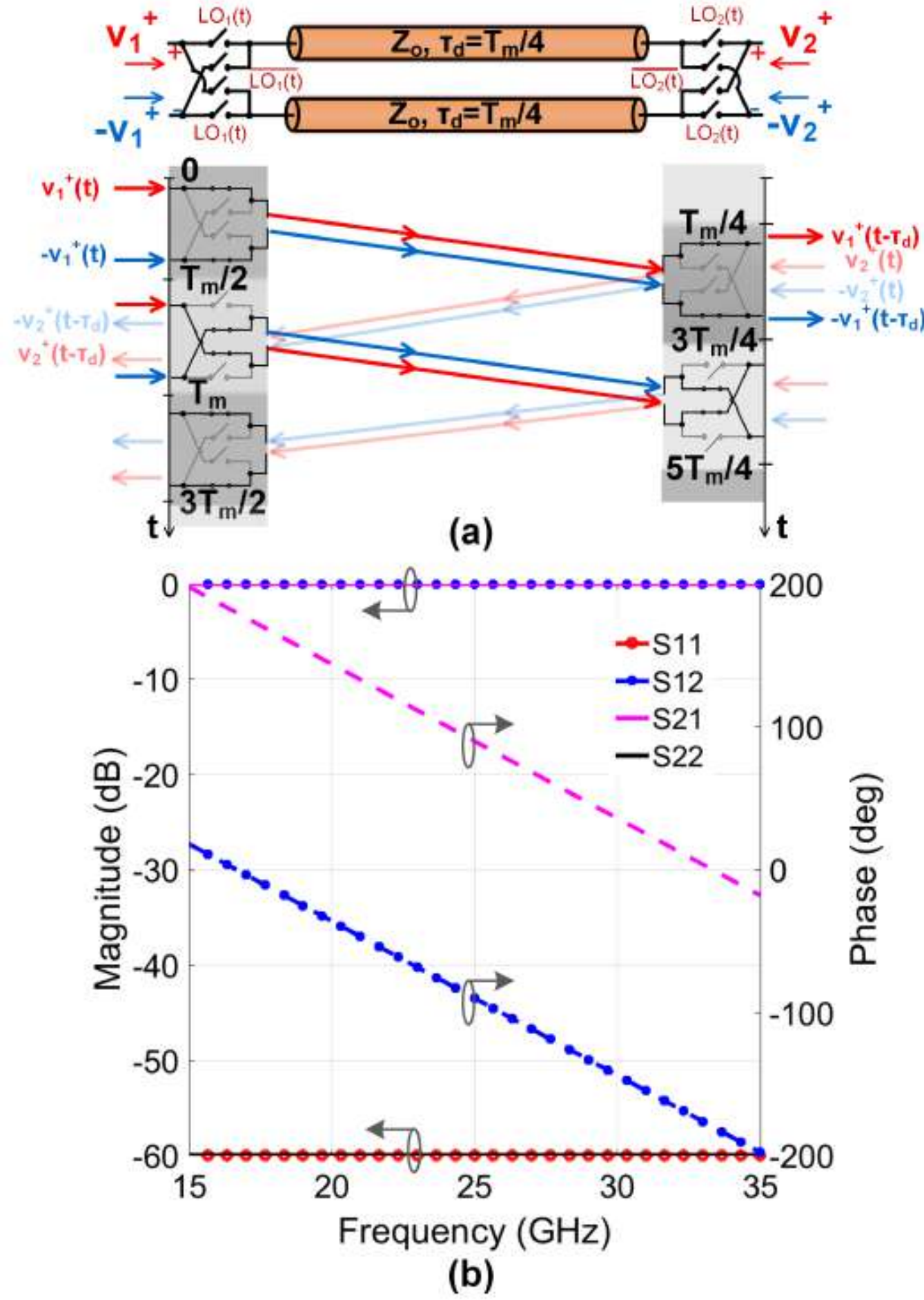

Fig. 3. Doubly-balanced ultra-broadband gyrator: (a) circuit diagram and operation, and (b) scattering parameters for $f_m$ = 8.33 GHz.

### D. Frequency-Domain Analysis of the Ultra-broadband Gyrator

The ultra-broadband gyrator can be interpreted as a transmission line sandwiched between two mixers, and can be analyzed in the frequency domain. The multiplication factors of the left-hand-side mixer and the right-hand-side mixer, $m(t)$ and $m(t-\frac{T_m}{4})$, can be expressed in the frequency-domain as shown in (10) and (11).

An input signal, $e^{j\omega t}$ traveling from left to right, after passing through the left-hand-side mixer, will be multiplied with the multiplication factor $m(t)$, resulting in (12). From this equation, we can see that the transmission line supports signals at infinitely-many intermodulation frequencies, and therefore, disperson-free operation of the line is critical. As can be

$$v_1^-(t) = m(t) \times v_2^+\left(t - \frac{T_m}{4}\right) \times m\left(t - \frac{T_m}{2}\right) = -v_2^+\left(t - \frac{T_m}{4}\right) \tag{7}$$

$$v_2^-(t) = m\left(t - \frac{T_m}{4}\right) \times v_1^+\left(t - \frac{T_m}{4}\right) \times m\left(t - \frac{T_m}{4}\right) = v_1^+\left(t - \frac{T_m}{4}\right) \tag{8}$$

$$m(t) = \frac{2}{j\pi}\sum_{n=1}^{\infty}\frac{1}{2n-1}(e^{j(2n-1)\omega_m t} - e^{-j(2n-1)\omega_m t}) \tag{10}$$

$$m(t - \frac{T_m}{4}) = \frac{2}{j\pi}\sum_{n=1}^{\infty}\frac{1}{2n-1}(e^{-j\frac{(2n-1)\pi}{2}}e^{j(2n-1)\omega_m t} - e^{j\frac{(2n-1)\pi}{2}}e^{-j(2n-1)\omega_m t}) \tag{11}$$

$$v_1^+(t) \times m(t) = \frac{2}{j\pi}\sum_{n=1}^{\infty}\frac{1}{2n-1}(e^{j(\omega+(2n-1)\omega_m)t} - e^{j(\omega-(2n-1)\omega_m)t}) \tag{12}$$

$$v_1^+(t - \frac{T_m}{4}) \times m(t - \frac{T_m}{4}) = \frac{2}{j\pi}e^{-j\omega\frac{T_m}{4}}\sum_{n=1}^{\infty}\frac{1}{2n-1}(e^{-j\frac{(2n-1)\pi}{2}}e^{j(\omega+(2n-1)\omega_m)t} - e^{j\frac{(2n-1)\pi}{2}}e^{j(\omega-(2n-1)\omega_m)t}) \tag{13}$$

$$\begin{aligned}
v_2^-(t) &= \left[\frac{2}{j\pi}e^{-j\omega\frac{T_m}{4}}\sum_{n=1}^{\infty}\frac{1}{2n-1}(e^{-j\frac{(2n-1)\pi}{2}}e^{j(\omega+(2n-1)\omega_m)t} - e^{j\frac{(2n-1)\pi}{2}}e^{j(\omega-(2n-1)\omega_m)t})\right] \\
&\times \left[\frac{2}{j\pi}\sum_{p=1}^{\infty}\frac{1}{2p-1}(e^{-j\frac{(2p-1)\pi}{2}}e^{j(2p-1)\omega_m t} - e^{j\frac{(2p-1)\pi}{2}}e^{-j(2p-1)\omega_m t})\right] \\
&= \frac{-4}{\pi^2}e^{-j\omega\frac{T_m}{4}}\sum_{n=1}^{\infty}\frac{1}{(2n-1)^2}(-e^{j(\omega+(2n-1)\omega_m-(2n-1)\omega_m)t} - e^{j(\omega-(2n-1)\omega_m+(2n-1)\omega_m)t}) \\
&= \frac{8}{\pi^2}e^{-j\omega\frac{T_m}{4}}e^{j\omega t}\sum_{n=1}^{\infty}\frac{1}{(2n-1)^2} \\
&= e^{-j\omega\frac{T_m}{4}}e^{j\omega t} \\
&= e^{-j\omega\frac{T_m}{4}}v_1^+(t)
\end{aligned} \tag{14}$$

seen, the amplitude of higher-order intermodulation products becomes progressively smaller, and therefore, dispersion at frequencies further away from the operation frequency will have a progressively smaller effect. A quantification of the impact of transmission-line dispersion is given in Section IV. This signal experiences a delay of $T_m/4$ through the transmission line, at which point, it can be expressed as shown in (13). This gets multiplied with the multiplication factor of the right-hand-side mixer, resulting in the output signal shown in (14). This agrees with the result from the time-domain analysis in (8). At frequencies other than $\omega$, due to cancellation of the intermodulation products, the amplitude of the signal becomes zero, making the structure appear externally LTI, while being internally LPTV, as mentioned earlier. For a signal travelling from right to left, the result in (7) can be obtained in the frequency domain as well by following a similar approach.

## III. Amplitude Non-reciprocity: Isolators and Circulators

So far, we have discussed various topologies of spatio-temporal conductivity-modulation that realize phase non-reciprocity. In this section, we discuss topologies and circuits exhibiting amplitude non-reciprocity, such as isolators and circulators.

### A. Frequency-conversion Isolator

An isolator is a two-port non-reciprocal component that supports transmission in one direction but not in the reverse direction. A frequency-conversion isolator can be realized from the doubly-balanced phase-non-reciprocal element by modifying the delay between the modulation signals of the left-hand-side Gilbert-quad switch set, $LO_1(t)$, and right-hand-side switch set, $LO_2(t)$, to a value $T_m/8$, which is also the modified propagation delay of the differential transmission line between the switch sets, as depicted in Fig. 4.

A signal travelling from left to right, incident differentially at port 1, transmits to port 2 with no sign inversion in the first half of the modulation cycle, and two inversions in the second half of the cycle, similar to the doubly-balanced phase-non-reciprocal element. A signal travelling from right to left, incident differentially at port 2, will be transmitted without sign inversion during the first and third quarter cycles. However, during the second and fourth quarter cycles, the signal will be transmitted with a sign inversion from the left and right Gilbert-quad switch sets respectively, as depicted in Fig. 4. Equivalently, the signal travelling from right to left will be multiplied with a $+1/-1$ square wave of $50\%$ duty cycle and $2\omega_m$ angular frequency, and gets up-/down-converted to other frequencies (i.e., inter-modulation products), leading to an isolation at the input frequency. The operation of this structure can be expressed in the time domain as

$$v_1^-(t) = m\left(t - \frac{T_m}{4}\right) \times v_2^+\left(t - \frac{T_m}{8}\right) \times m(t) \tag{15}$$

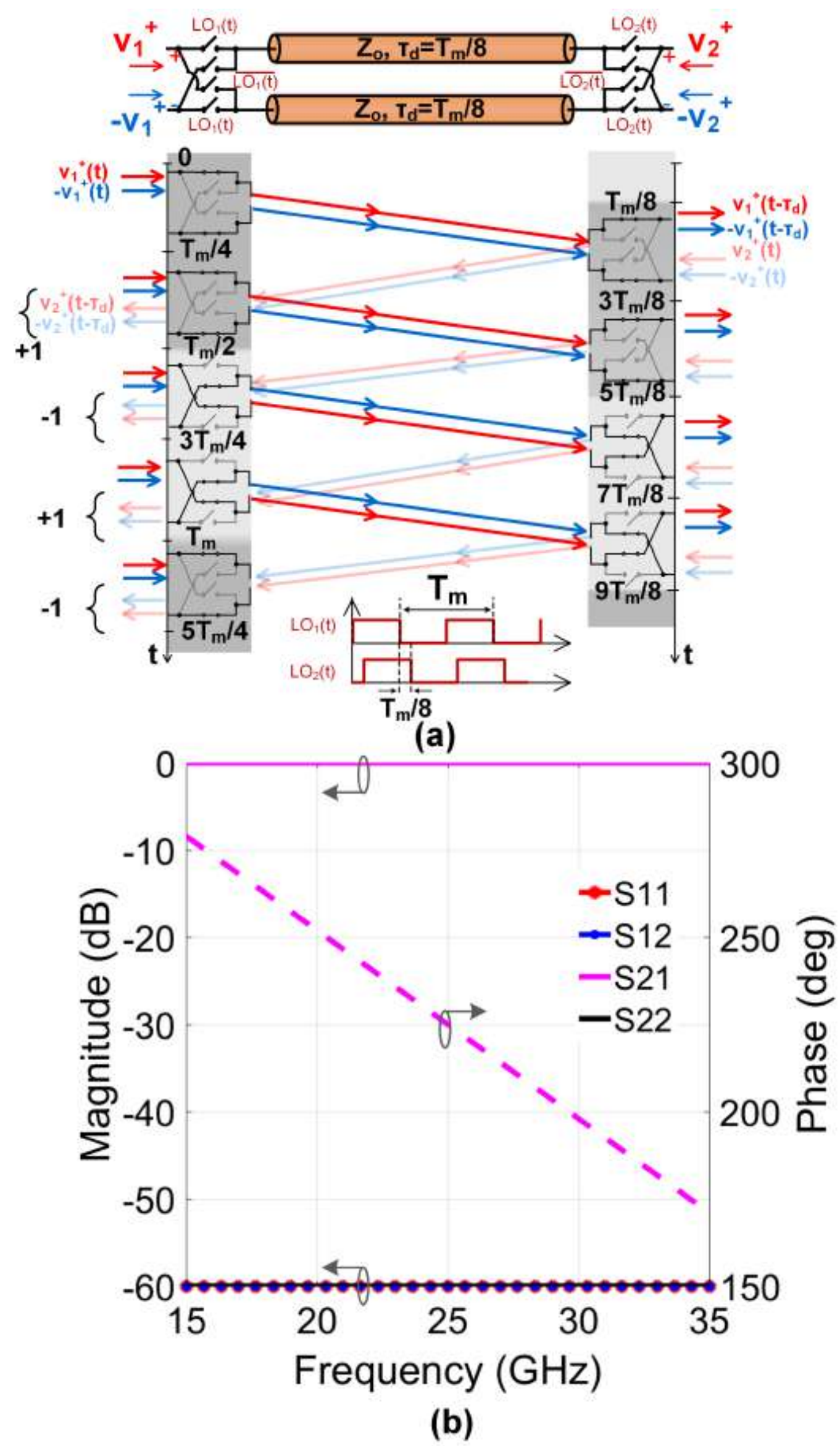

Fig. 4. Frequency-conversion isolator: (a) circuit diagram and operation, and (b) fundamental-to-fundamental scattering parameters for $f_m$ = 8.33 GHz.

$$
\begin{aligned}
v_2^-(t) &= m\left(t-\frac{T_m}{8}\right)\times v_1^+\left(t-\frac{T_m}{8}\right)\times m\left(t-\frac{T_m}{8}\right) \\
&= v_1^+\left(t-\frac{T_m}{8}\right)
\end{aligned}
\quad . \tag{16}
$$

The fundamental-to-fundamental S-parameter matrix is obtained by taking Fourier transform of the time-domain equations:

$$
S=\begin{pmatrix} 0 & 0 \\ +e^{-j\omega\frac{T_m}{8}} & 0 \end{pmatrix} \tag{17}
$$

From (17), it can be seen that the power incident at port 2 at any input frequency gets translated to other frequencies, while the power incident at port 1 is transmitted losslessly, making it a frequency-conversion isolator that can operate over an infinitely-wide range of operating frequencies. It should be emphasized, however, that this configuration has limitations when it comes to instantaneously wideband signals at port 2. For signals with bandwidth greater $2\omega_m$, the frequency-translated signals will fall within the desired signal's bandwidth, compromising the isolation from port 2 to port 1. Fig. 4(b) shows the fundamental-to-fundamental S parameters around 25 GHz for $f_m$=8.33 GHz.

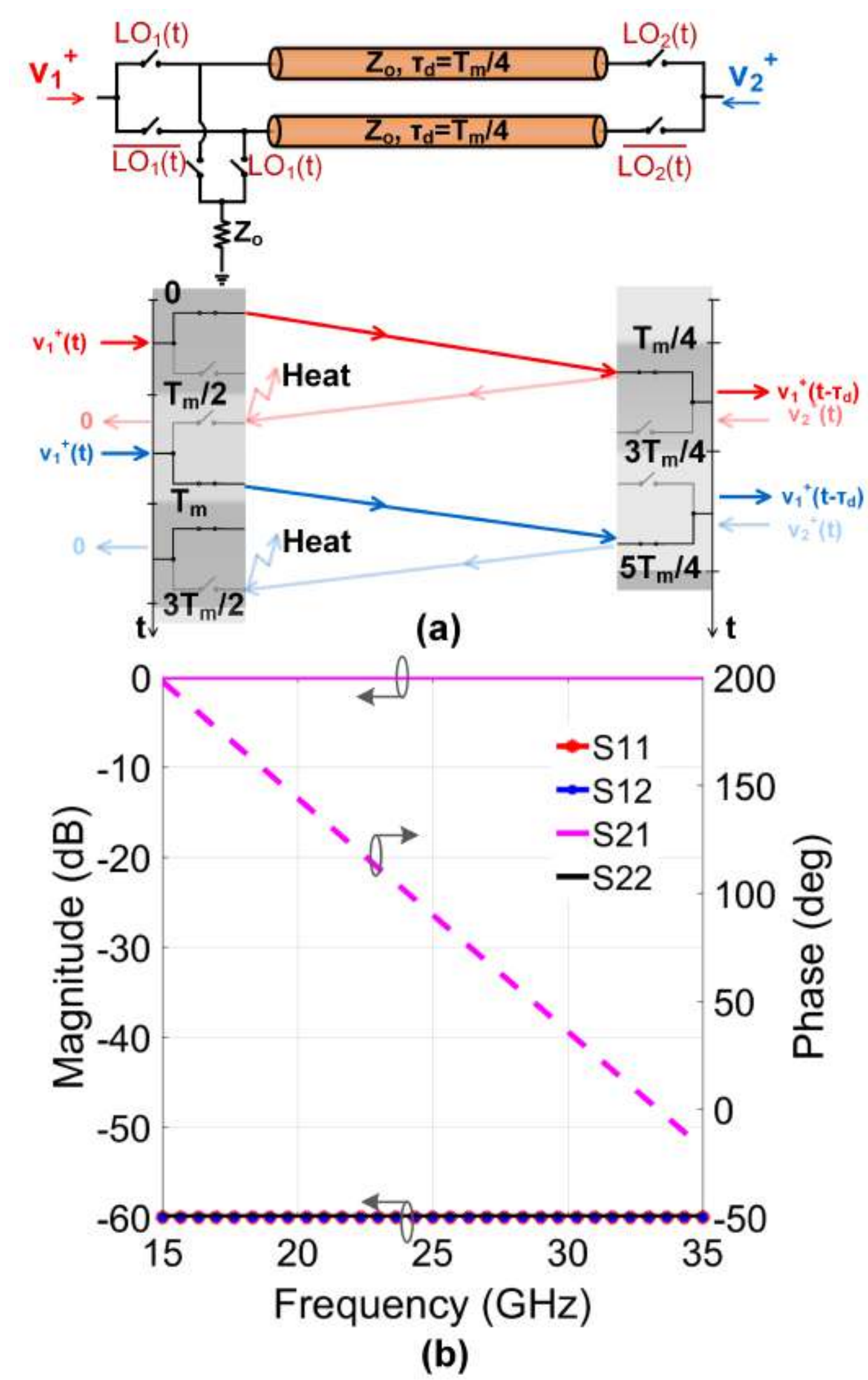

Fig. 5. Ultra-broadband dissipative isolator: (a) circuit diagram and operation, and (b) scattering parameters for the case with $f_m$ = 8.33 GHz.

### B. Ultra-broadband Dissipative Isolator

Isolation through frequency conversion is often undesirable, as isolators are typically used to protect sources and amplifiers from back reflections. In such situations, it is desirable for signals incident in the reverse direction to be dissipated within the isolator. Indeed, such perfect isolation and matching cannot be realized without the presence of loss in the system [39]. An ultra-broadband dissipative isolator with no frequency conversion can be realized by modifying the balanced phase-non-reciprocal element by adding a pair of parallel switches with a terminating resistor, of value equal to the port impedance, at the left-hand-side switches. The new pair of switches is modulated using the complementary clocks of the left-hand-side switches, as shown in Fig. 5. For signals travelling from left to right, this structure behaves exactly like the balanced phase-non-reciprocal element, imparting a delay of $T_m/4$. However, for signals travelling from right to left, after the first pass, instead of reflecting, the signals travel through the newly-added switches and get dissipated in the terminating resistor. The operation of the structure can be expressed in the time domain as

$$v_1^-(t) = 0, \tag{18}$$

$$v_2^-(t) = v_1^+\left(t - \frac{T_m}{4}\right). \tag{19}$$

By taking Fourier transform of the time-domain equations, the S-parameter matrix can be calculated as

$$S = \begin{pmatrix} 0 & 0 \\ +e^{-j\omega\frac{T_m}{4}} & 0 \end{pmatrix}. \tag{20}$$

Theoretically, this structure appears externally-LTI, exhibits lossless transmission in the forward direction, perfect isolation in the reverse direction, and has infinite bandwidth (Fig. 5(b)). In practice, the loss and bandwidth are limited by quality factor of transmission line, non-zero switch ON resistance, dispersion of the transmission line, switch parasitic capacitance, and non-idealities of the modulation clocks, similar to the phase-non-reciprocal elements.

### C. Ultra-broadband Circulator

The ultra-broadband dissipative isolator is actually a special case of an ultra-broadband circulator whose third port has been terminated to the reference impedance. In other words, a circulator with infinite bandwidth can be realized by replacing the terminating resistor in the ultra-broadband isolator by a third port as shown in Fig. 6. For such a circulator, it can be shown using time domain analysis that lossless transmission happens from port 1 to port 2 , port 2 to port 3 and port 3 to port 1, and there will be a perfect isolation in the reverse circulation direction. The S-parameter matrix for this ultra-broadband circulator can be constructed as

$$S = \begin{pmatrix} 0 & 0 & e^{-j\omega\frac{2T_m}{4}} \\ e^{-j\omega\frac{T_m}{4}} & 0 & 0 \\ 0 & e^{-j\omega\frac{T_m}{4}} & 0 \end{pmatrix}. \tag{21}$$

As opposed to three switches in the signal path as in [32] this structure has only two switches, enabling it to potentially have superior insertion loss performance. In addition, this structure needs only two transmission lines as opposed to six transmission lines in [32], making it more compact and readily implementable at RF frequencies. It is closely related to the 4-port circulator recently described in [33].

### D. High-linearity Circulator

The ultra-broadband circulator features switches in the signal path, which would limit the power handling of the circulator, particularly in the TX-to-ANT path. Recently, we demonstrated a circulator architecture with high power handling by wrapping a $3\lambda/4$ transmission line with a characteristic impedance equal to the port impedance around the gyrator component with $\pm 90°$ non-reciprocal phase, and placing ports $\lambda/4$ apart on the $3\lambda/4$ transmission line [29]–[31] (Fig. 7). For such a circulator, the S parameters at the center frequency can be computed as

$$S = \begin{pmatrix} 0 & 0 & -1 \\ -j & 0 & 0 \\ 0 & -j & 0 \end{pmatrix} \tag{22}$$

The power handling at port 1 of the circulator (TX port) can be improved significantly by placing the third port at the gyrator element ($l$=0 in Fig. 7), as the isolation from port 1 to port 3 suppresses the voltage swing across the gyrator [29]. In [29], a 750MHz circulator was demonstrated in 65nm CMOS using the N-path-filter-based gyrator discussed earlier. In [30], [31], a differential circulator was realized at 25GHz in 45nm SOI CMOS using the doubly-balanced transmission-line based gyrator. In general, both single-ended and differential circulators can be realized using the balanced and doubly-balanced gyrator configurations (Figs. 7(a) and (b)). The balanced configuration achieves gyrator functionality at $\omega = (2n+1)\omega_m$. The doubly-balanced configuration achieves a non-reciprocal phase of $\pi$ across infinite bandwidth, but the need for $\pm 90°$ phase once again dictates $\omega = (2n+1)\omega_m$. In the 25GHz circulator in [30], [31], the modulation frequency was one-third of the operating frequency, which critically enabled scaling of the circulator to millimeter-waves. An important point of difference is that the internal gyrator nodes also experience suppressed voltage swing for TX-port excitations in the differential configuration, due to the fact that a pair of switches is always on, shorting the internal nodes to the external nodes (Fig. 7(c)). This, however, is not the case for the single-ended circulator configuration that uses the balanced gyrator, limiting the TX power handling enhancement. This can be restored by adding a pair of parallel switches to ground at both the left-hand side and right-hand side switches. The new pair of switches would be modulated using the complementary clocks of the original switches, making the nodes $SB_{n2}$ and $SB_{n3}$ quiet. However, due to the additional switches, the power consumption of this single-ended circulator would become equal to that of the differential circulator, and so it is beneficial to implement the differential configuration owing to its benefits of lower clock feedthrough, and +3dB higher power handling.

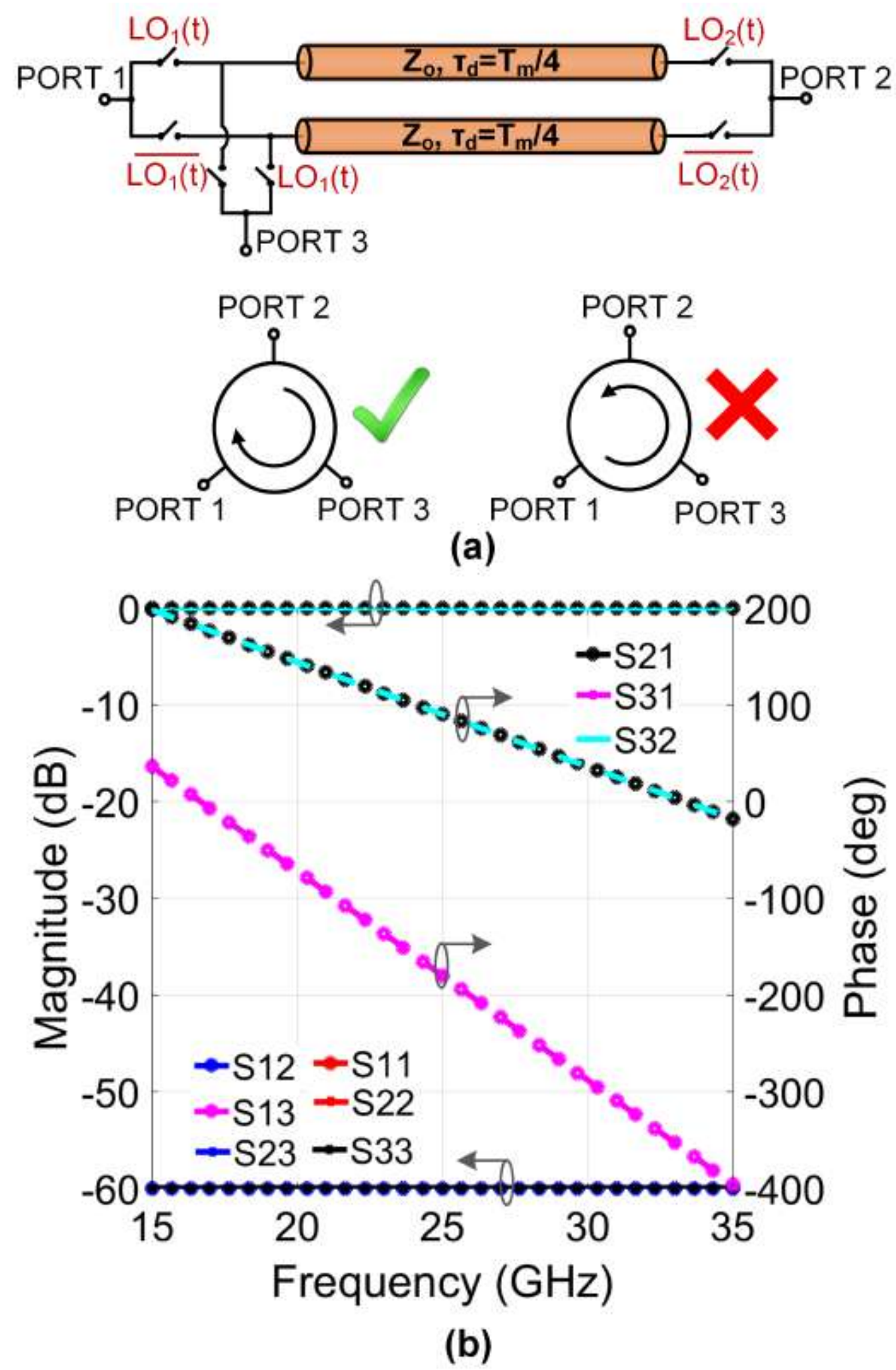

Fig. 6. Ultra-broadband circulator: (a) circuit diagram, and (b) scattering parameters for $f_m$ = 8.33 GHz.

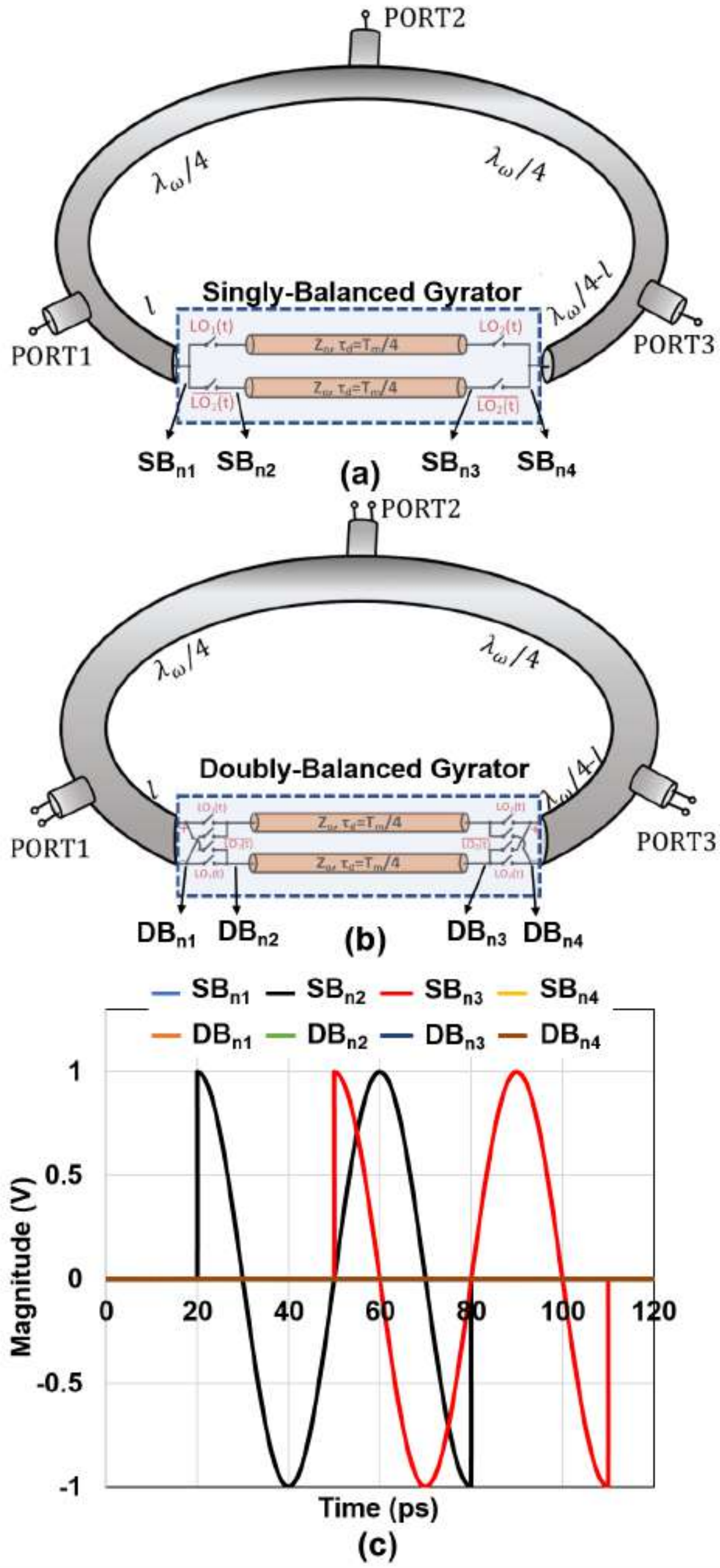

Fig. 7. Highly-linear circulators: (a) single-ended configuration leveraging the balanced gyrator, and (b) differential configuration leveraging the doubly-balanced gyrator. (c) Internal node voltage swings of both configurations when l=0 for 1V TX port voltage at 25GHz.

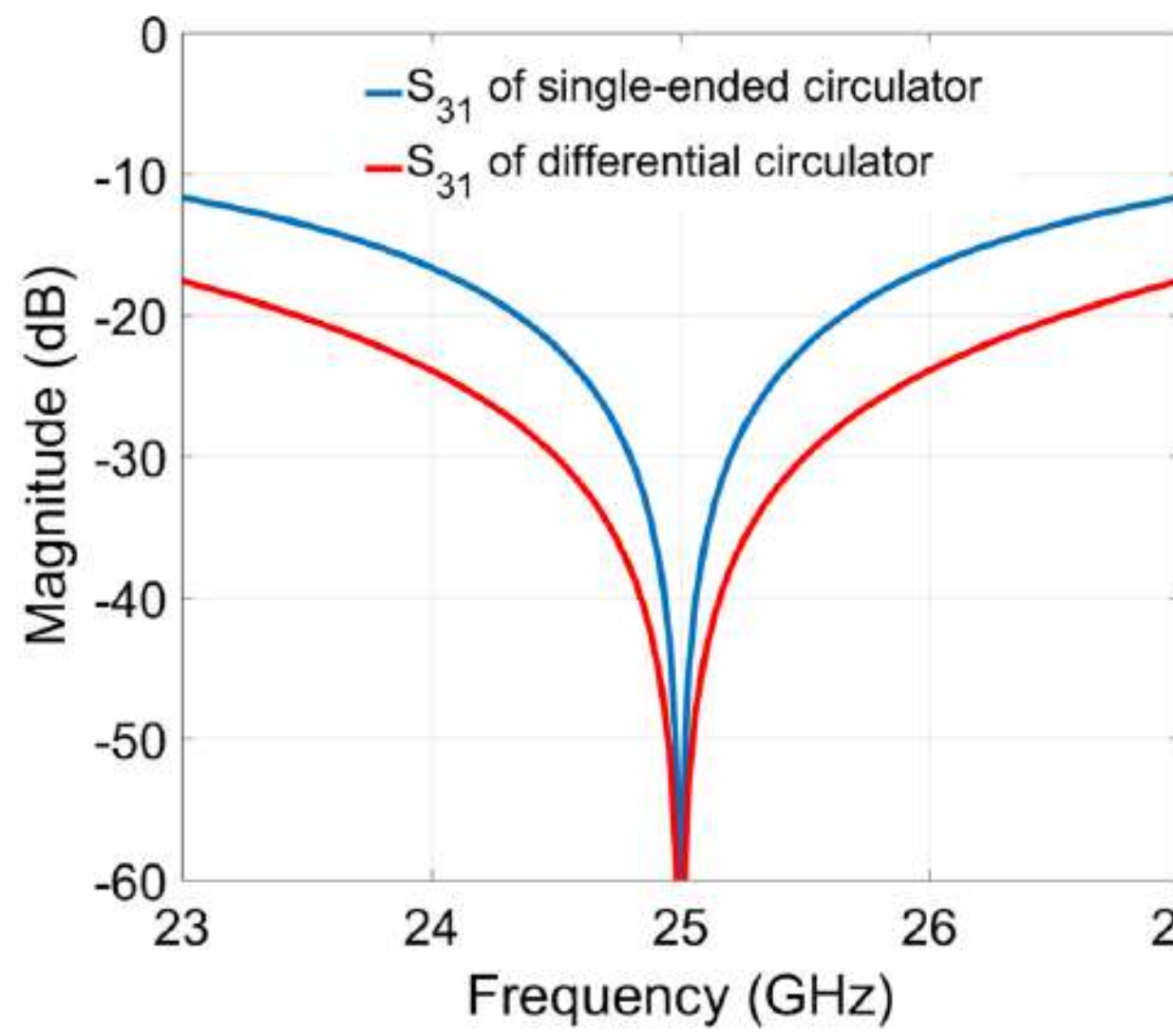

Fig. 8. Port 1-Port 3 isolation of the differential circulator featuring the doubly-balanced gyrator and the single-ended circulator featuring the balanced gyrator, depicted in Figs.7(a) and (b).

Isolation from port 1 to port 3 is due to destructive interference of signals from $\lambda/2$ reciprocal path and the $\lambda/4$ transmission line and gyrator path. Perfect isolation occurs at the center frequency where the signals from both the paths are exactly out of phase. For frequencies off center, the condition of destructive interference starts to fail and isolation will be limited to a finite value. Hence, the isolation bandwidth is dependent on the phase response of the $3\lambda/4$ ring and the non-reciprocal phase response of the gyrator. The doubly-balanced gyrator exhibits a non-reciprocal phase response of 180° over theoretically infinite bandwidth, while the balanced phase-non-reciprocal element behaves as a gyrator only at discrete frequencies, as discussed earlier. Hence, as shown in Fig. 8, the 20dB isolation bandwidth of the differential circulator featuring the doubly-balanced gyrator is 2.34× higher than that of the single-ended circulator featuring the balanced gyrator.

## IV. Analysis of Insertion Loss Based on Various Non-Idealities

In this section of the paper, we derive formulae for estimating the transmission losses of the highly-linear circulator architecture taking various non-idealities into account using perturbation analysis. During implementation, transmission lines of length comparable to the wavelength are often miniaturized using lumped L-C components to reduce the chip area and to absorb the parasitic capacitances of the switches into the transmission lines. Hence, in our analysis, it is assumed that the $\lambda/4$ transmission lines sections in the $3\lambda/4$ ring were miniaturized using C-L-C sections. Various non-idealities such as quality factor of the transmission lines/inductors, non-zero switch resistance of the switches in the gyrator element, rise/fall time of the modulation signals, and the dispersion of the transmission line in the gyrator, limit the insertion loss of the circulator. In this derivation, it has been assumed that the effect of the non-idealities are small, so that they do not degrade the matching of the gyrator element and the isolation of the circulator substantially. The effect of each non-ideality in the circulator is separately computed using perturbation analysis, and the total loss of the circulator is estimated by summing up the individual contributions.

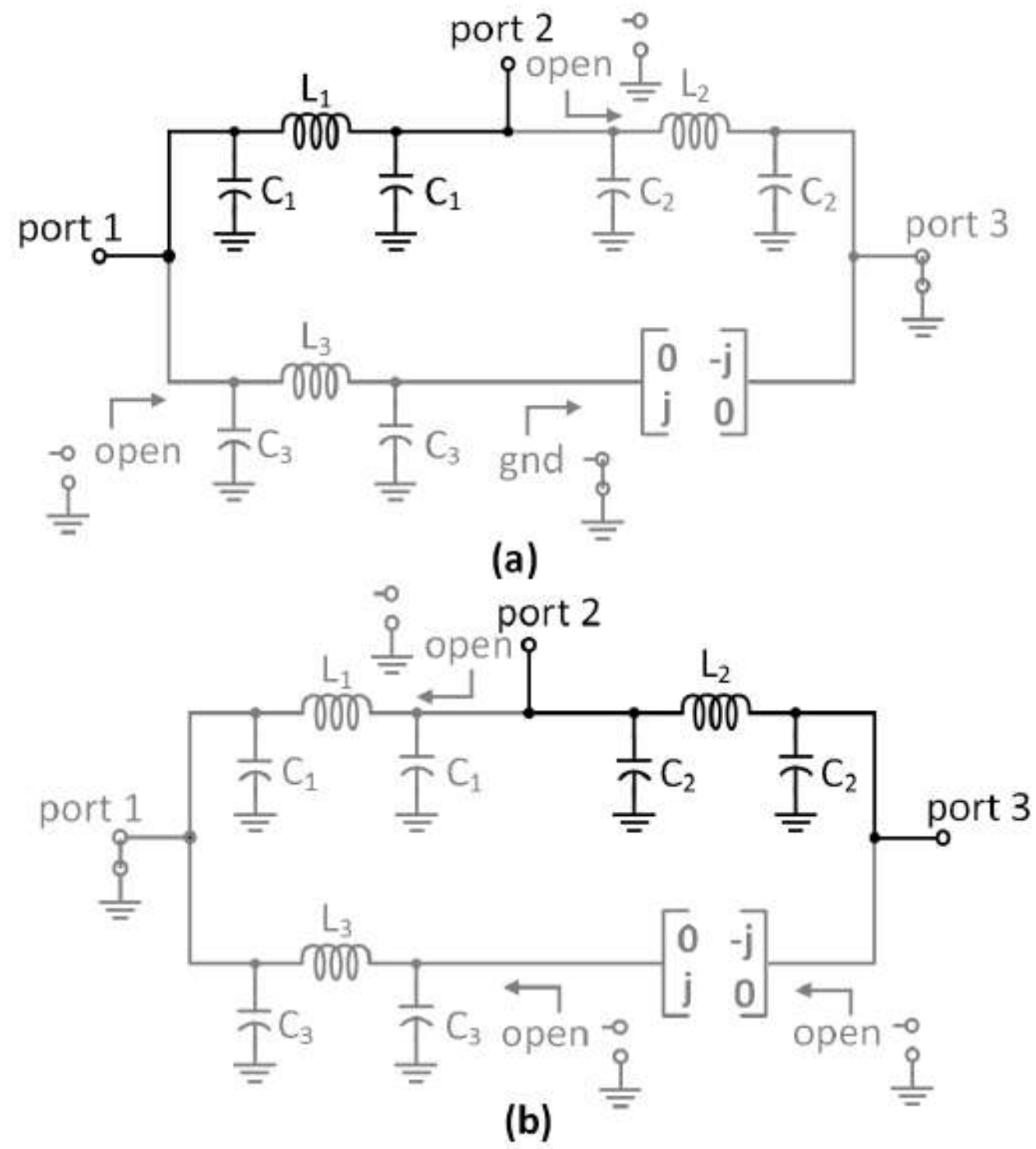

Fig. 9. (a) Equivalent circuit for loss estimation from port 1 to port 2. (b) Equivalent circuit for loss estimation from port 2 to port 3.

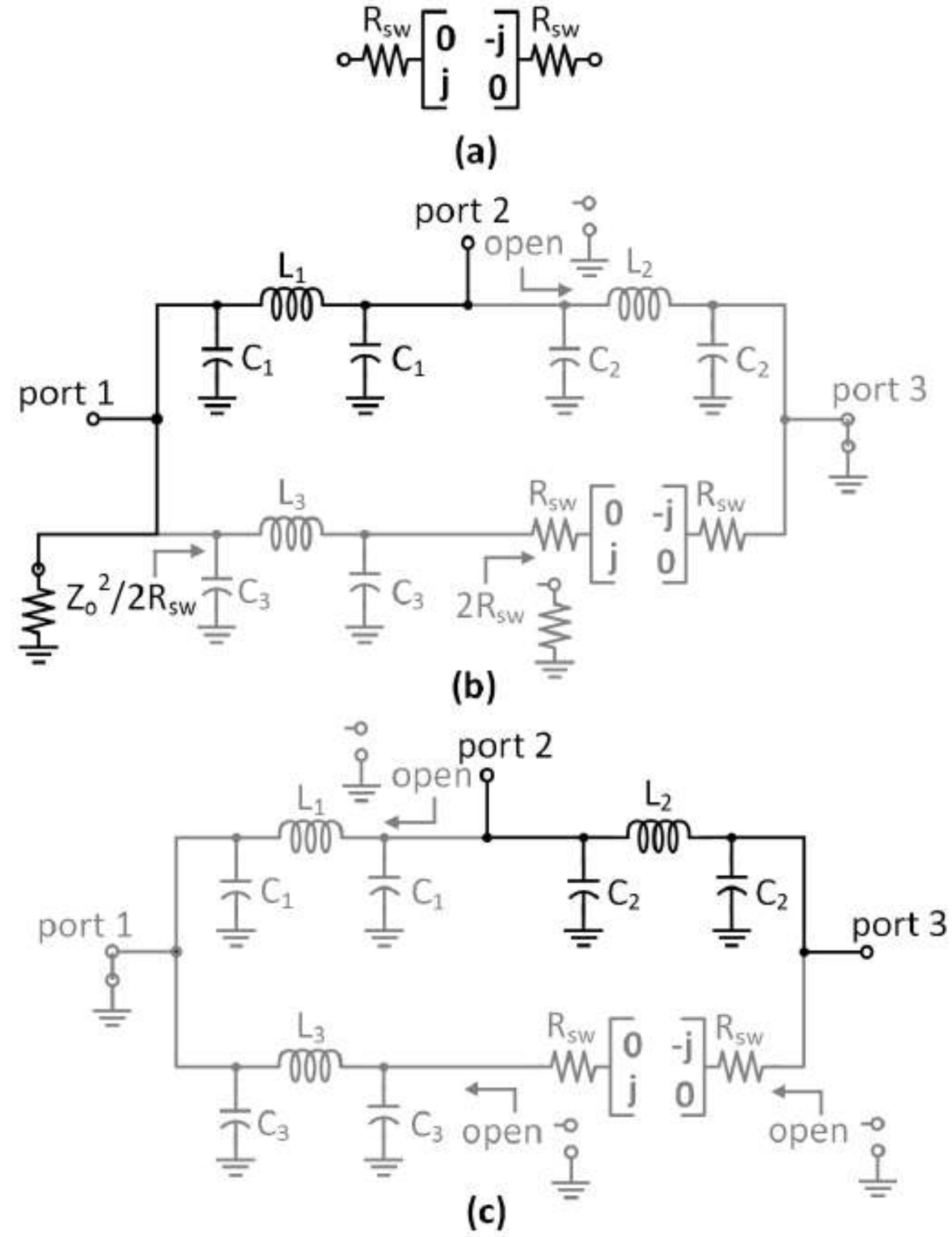

Fig. 10. (a) Equivalent circuit for a gyrator with non-zero switch resistance. (b) Equivalent circuit for loss estimation from port 1 to port 2 of the circulator with non-zero switch resistance. (b)Equivalent circuit for loss estimation from port 2 to port 3 of the circulator with non-zero switch resistance.

While estimating the loss from port 1 to port 2 in the circulator, port 3 can be shorted to ground due to the port 1-to-port 3 isolation. As a result, the $\lambda/4$ section between port 2 and port 3 will transform the short circuit to an open circuit when seen from port 2 as will the the $\lambda/4$ section between port 1 and gyrator when seen from port 1. Hence, the circulator reduces to a $\lambda/4$ transmission line connecting ports 1 and 2 as shown in Fig. 9(a). A similar equivalent circuit for port 2-to-port 3 transmission is shown in Fig. 9(b).

### A. Effect of Non-zero Switch Resistance, $R_{sw}$

A switch with non-zero ON resistance, $R_{sw}$, can be modeled by an ideal switch with a series resistance, $R_{sw}$. Hence, a gyrator with non-zero switch resistance can be modeled in the manner depicted in Fig. 10(a). As before, port 3 of the circulator can be shorted to ground when estimating the loss from port 1 to port 2, as depicted in Fig. 10(b). When one of its ports is shorted to ground, the input impedance of the gyrator shown in Fig. 10(a) at the other port is $2R_{sw}$. This $2R_{sw}$ resistance will be transformed to $Z_0^2/2R_{sw}$ at port 1 as depicted in Fig. 10(b). Consequently, the transmission loss from port 1 to port 2 can be written as

$$|S_{21}| \approx \frac{Z_0}{Z_0 + R_{sw}} = \frac{Q_{sw}}{Q_{sw}+1}. \tag{23}$$

where $Q_{sw}$, the *equivalent* quality factor of the switch, is given by $Z_0/R_{sw}$. Similarly, port 1 can be shorted to ground while estimating transmission loss from port 2 to port 3. Using similar arguments, the circulator can be reduced to the circuit depicted in Fig. 10(c). Interestingly, the non-zero switch resistance will have no effect on the transmisison loss from port 2 to port 3 because the switch resistance is in series with an open termination.

$$|S_{32}| \approx 1. \tag{24}$$

### B. Effect of the Quality Factor, $Q_1$, of Inductor $L_1$

Similar arguments can be made for the quality factor, $Q_1$, of inductor $L_1$. Figs. 11 (a) and (b) depict the equivalent circuits for $S_{21}$ and $S_{32}$ loss estimation. We have assumed that the inductor losses are purely in the form of series resistance in writing the equivalent parallel resistance as $\omega L_1 Q_1 = Q_1 Z_0$.

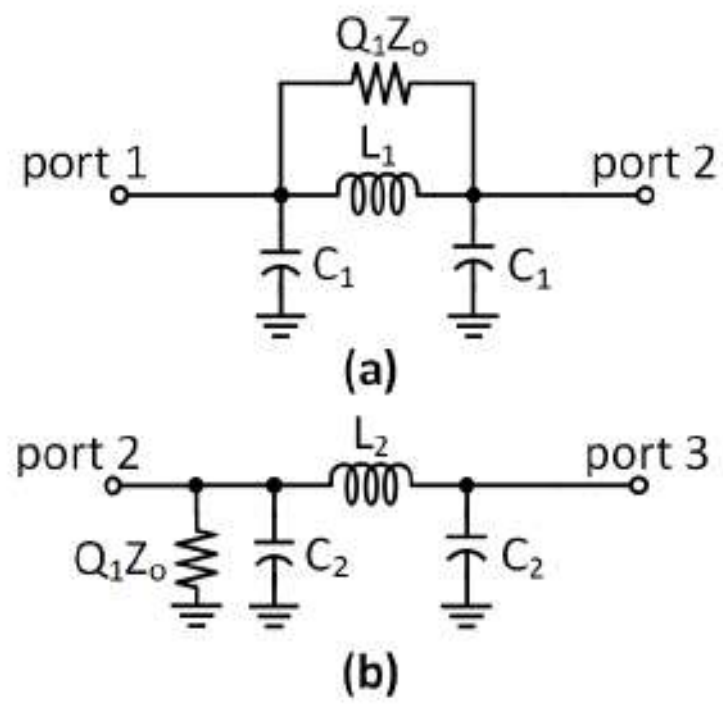

Fig. 11. Equivalent circuits for loss estimation of the circulator with finite quality factor, $Q_1$, in inductor $L_1$: (a) port 1 to port 2, and (b) port 2 to port 3.

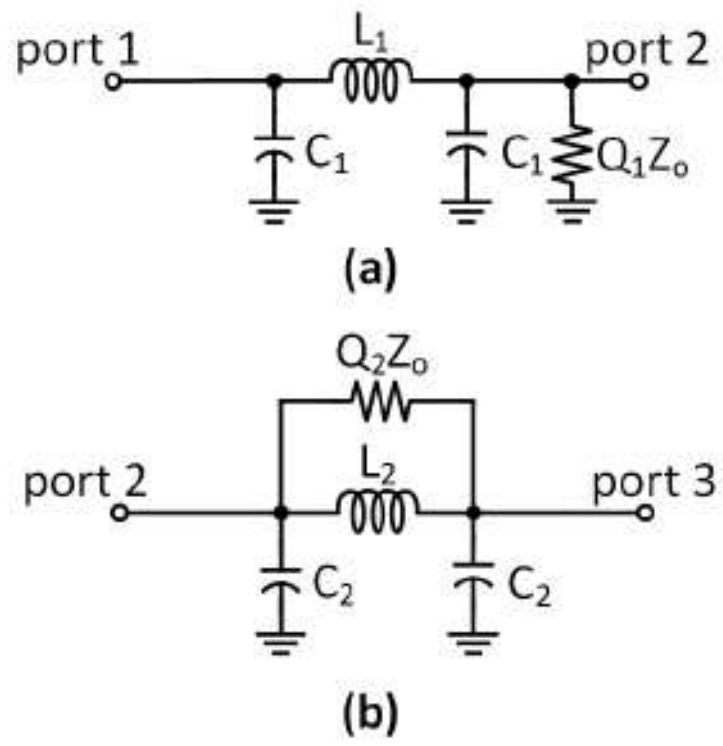

Fig. 12. (a) Equivalent circuit for loss estimation from port 1 to port 2 of the circulator with inductor $L_2$ of finite quality factor, $Q_2$. (b)Equivalent circuit for loss estimation from port 2 to port 3 of the circulator with inductor $L_2$ of finite quality factor, $Q_2$.

Further, we have assumed that the capacitors used to lump the transmission line are lossless. By analyzing these circuits, we can write the transmission losses as

$$|S_{21}| \approx |\frac{(\frac{1}{Q_1} - j)}{(1 + \frac{1}{Q_1}) + \frac{j}{Q_1}}| \approx \frac{Q_1}{Q_1 + 1}, \tag{25}$$

$$|S_{32}| \approx \frac{2Q_1}{2Q_1 + 1}. \tag{26}$$

### C. Effect of the Quality Factor, $Q_2$, of Inductor $L_2$

Making similar arguments, the equivalent circuits in Fig. 12 enable loss estimation in the presence of finite quality factor $Q_2$ in inductor $L_2$. By analyzing these circuits, we can write the transmission losses as

$$|S_{21}| \approx \frac{2Q_2}{2Q_2 + 1}, \tag{27}$$

$$|S_{32}| \approx |\frac{(\frac{1}{Q_2} - j)}{(1 + \frac{1}{Q_2}) + \frac{j}{Q_2}}| \approx \frac{Q_2}{Q_2 + 1}. \tag{28}$$

### D. Effect of the Quality Factor, $Q_3$, of Inductor $L_3$

Similarly, the circulator with finite quality factor, $Q_3$, in inductor $L_3$ can be reduced to a C-L-C section with a shunt resistor, $Q_3Z_0$, at port 1 while estimating loss from port 1 to port 2 and a C-L-C section with a shunt resistor, $Q_3Z_0$, at port 3 while estimating loss from port 2 to port 3. By analyzing these circuits, we can write the transmission losses as

$$|S_{21}| \approx \frac{2Q_3}{2Q_3 + 1}, \tag{29}$$

$$|S_{32}| \approx \frac{2Q_3}{2Q_3 + 1}. \tag{30}$$

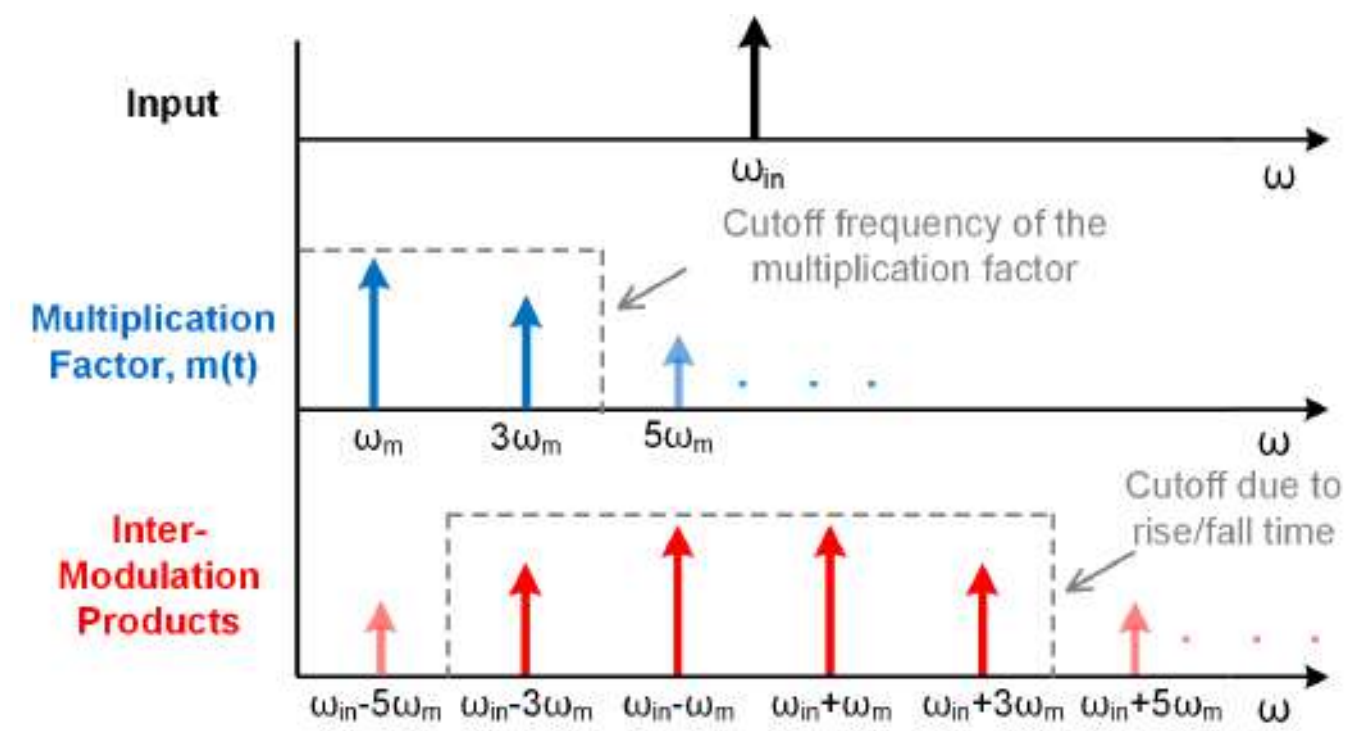

Fig. 13. Filtering due to finite rise/fall time of the modulation signals.

### E. Effect of the Rise/Fall Time of the Modulation Signal

In Section II-D, we have presented a frequency-domain of the doubly-balanced ultra-broadband gyrator with the switch sets viewed as mixers. We considered the modulation signals to be perfect square waves with negligible rise/fall time, but in a practical implementation, the rise/fall time ($t_r/t_f$) can be a significant portion of the time period, particularly for high-frequency operation. For a trapezoidal modulation signal with rise/fall time of $t_r = t_f$, the frequency content of modulation signal will be limited to a cutoff frequency of $(0.35/t_r)\,Hz$. This will correspondingly establish an upper limit on the frequency content of the multiplication factor, $m(t)$, as shown in Fig. 13. The value of this cutoff frequency can be obtained from simulations. The effect of this cutoff frequency in the multiplication factor is a limit to the intermodulation products that are generated and travel down the line, as depicted in Fig. 13, resulting in loss of signal power. In our analysis, the filtering profile of this effect is assumed to be a brick wall, and mathematically, this can be expressed as a truncation in the higher frequency terms in (10) - 14. For the example shown in Fig. 13, where the cutoff frequency is between $3\omega_m$ and $5\omega_m$, 14 can be modified as

$$\begin{aligned} v_2^-(t) &= \frac{-4}{\pi^2} e^{-j\omega\frac{T_m}{4}} [(-e^{j(\omega+\omega_m-\omega_m)t} - e^{j(\omega-\omega_m+\omega_m)t}) \\ &\quad + \frac{1}{3^2}(-e^{j(\omega+3\omega_m-3\omega_m)t} - e^{j(\omega-3\omega_m+3\omega_m)t})] \\ &= \frac{80}{9\pi^2} e^{-j\omega\frac{T_m}{4}} e^{j\omega t} \\ &= \frac{80}{9\pi^2} e^{-j\omega\frac{T_m}{4}} v_1^+(t) \\ &\approx 0.9 e^{-j\omega\frac{T_m}{4}} v_1^+(t). \end{aligned} \tag{31}$$

Similarly, for the signal traveling from right to left, the signal transmitted can be written as

$$v_1^-(t) \approx -0.9e^{-j\omega\frac{T_m}{4}}v_1^+(t). \tag{32}$$

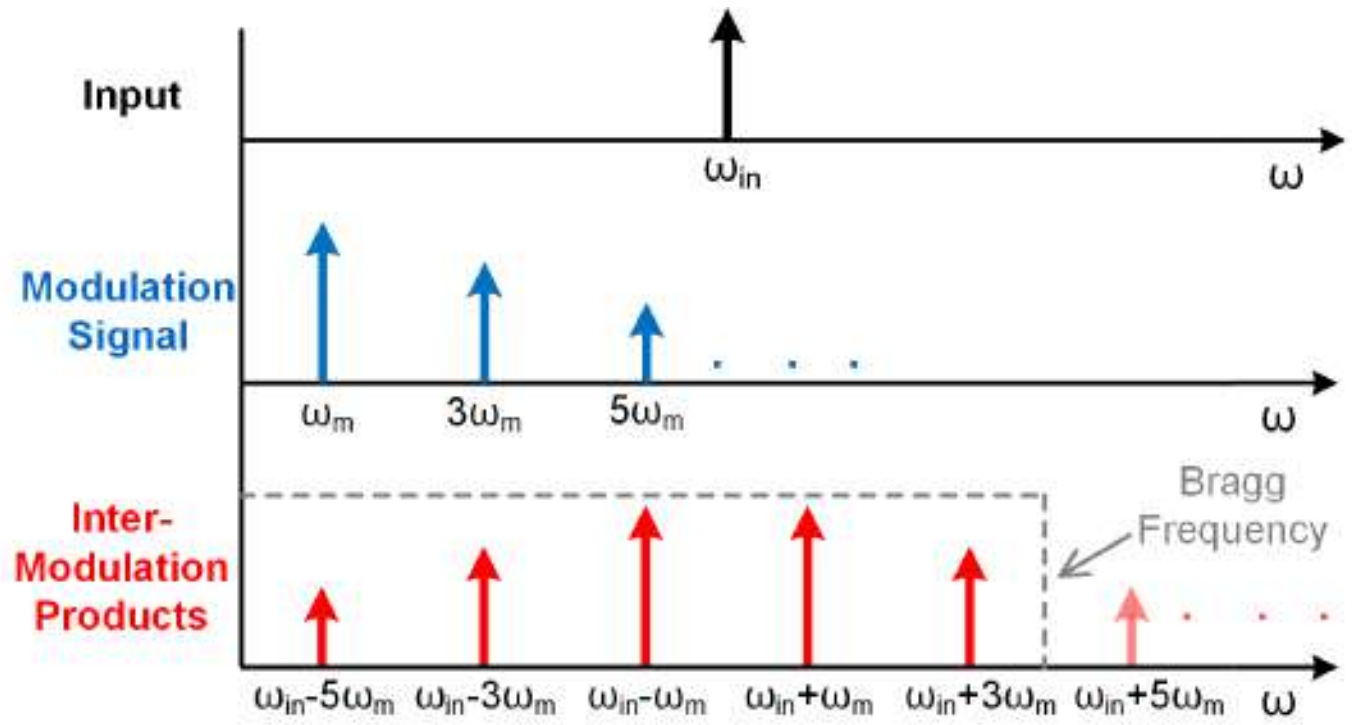

Fig. 14. Filtering of the inter-modulation products higher than the Bragg Frequency.

Assuming that the matching of the gyrator has not degraded due to the finite rise and fall time, the S-parameter matrix of the gyrator at the center frequency can be written as

$$S = \begin{pmatrix} 0 & k_{rise/fall}e^{j\frac{\pi}{2}} \\ k_{rise/fall}e^{-j\frac{\pi}{2}} & 0 \end{pmatrix} \tag{33}$$

where $k_{rise/fall}$ is the attenuation factor due to the filtering of higher inter-modulation products. For the example shown in Fig. 13, $k_{rise/fall} \approx 0.9$. Using the perturbation analysis discussed before, the S-parameters of the circulator can be expressed as

$$|S_{21}| \approx \frac{Q_{rise/fall}}{Q_{rise/fall}+1}, \tag{34}$$

$$|S_{32}| \approx \frac{Q_{rise/fall}}{Q_{rise/fall}+1}, \tag{35}$$

where $Q_{rise/fall}$ is an *effective* quality factor associated with the filtering due to finite rise/fall time of the modulation signals, and is given by $2(\frac{1+k_{rise/fall}^2}{1-k_{rise/fall}^2})$.

### *F. Effect of the Bragg Frequency of the Gyrator Transmission Line*

From the frequency-domain analysis, it is clear that the transmission line(s) between the switches must be able to support all the inter-modulation frequencies with the same group delay for a perfect lossless gyrator. If the transmission lines are miniaturized through a quasi-distributed implementation consisting of periodic lumped-LC sections, then the lines will have a cutoff frequency, after which the periodic structure no longer supports a traveling wave. This cutoff frequency is known as the Bragg frequency. Inter-modulation products above the Bragg frequency will be filtered and the signal power associated with them will be lost as depicted in Fig. 14. Similar to effect of finite rise and fall time of the modulation signal, the filtering profile due to Bragg frequency is also assumed to be a brick wall. All the inter-modulation terms with frequency above the Bragg frequency should be truncated from (14), and an associated attenuation factor, $k_{Bragg}$ can be calculated. Assuming that the matching of the gyrator has not degraded due to Bragg frequency effects, the S-parameter matrix of the gyrator at the center frequency can be written as

$$S = \begin{pmatrix} 0 & k_{Bragg}e^{j\frac{\pi}{2}} \\ k_{Bragg}e^{-j\frac{\pi}{2}} & 0 \end{pmatrix}. \tag{36}$$

From a similar perturbation analysis, the S-parameters of the circulator can be expressed as

$$|S_{21}| \approx \frac{Q_{Bragg}}{Q_{Bragg}+1} \tag{37}$$

$$|S_{32}| \approx \frac{Q_{Bragg}}{Q_{Bragg}+1} \tag{38}$$

where $Q_{Bragg}$ is an effective quality factor associated with the Bragg effect, and is given by $2(\frac{1+k_{Bragg}^2}{1-k_{Bragg}^2})$.

Both finite rise/fall time and finite Bragg frequency result in suppressing the higher frequency content of the inter-modulation products. Hence, one has to be careful while truncating the terms in (14) because the minimum of these cutoff frequencies will define the actual cutoff frequency for the inter-modulation products. In the final loss estimation, to avoid repetition of the loss due to filtering of higher modulation products, a general attenuation factor $k_{filter}$ can be calculated by truncating all the terms which are required to be truncated from Bragg frequency effects and finite rise/fall time effects. In general, for circulators operating at high frequencies, finite rise/fall time effects will dominate, while for circulators operating at lower frequencies, the need to aggressively miniaturize the transmission lines will cause Bragg frequency effects to dominate. For instance, in the 25GHz 45nm SOI CMOS circulator case study presented later in this paper, truncation was limited by the rise/fall time of the modulation signal, making $k_{filter} = k_{rise/fall}$.

### *G. Effect of the Quality Factor, $Q_{NR}$, of the Gyrator Transmission Line*

Due to ohmic losses, the signal attenuates exponentially as it travels through the transmission line in the gyrator, creating another loss mechanism. As mentioned earlier, all inter-modulation products travel through the transmission line, and the attenuation suffered by each inter-modulation product is different because attenuation constant depends on frequency. The attenuation suffered by a signal at a frequency $\omega$ as it travels through the transmission line with a delay of $T_m/4$ is $e^{-\frac{\pi}{4Q}\frac{\omega}{\omega_m}}$, where $Q$ is the quality factor of the transmission line at $\omega$. Due to the skin effect, $\rho \propto \sqrt{\omega}$, where $\rho$ is the resistivity of the metal. A general attenuation factor can be written by taking the skin effect of the metal into consideration as $e^{-\frac{(2n+1)\pi}{4Q_{NR}}\sqrt{\frac{\omega}{(2n+1)\omega_m}}}$, where $Q_{NR}$ is the quality factor of the transmission line at a frequency $\omega = (2n+1)\omega_m$. By modifying (13), the signal after traveling through the transmission line can be expressed as shown in (39). This signal gets multiplied with the multiplication factor of the right-hand-side mixer. Similar to (14), the signal voltage at the input frequency, $\omega$, at the other port of the gyrator can be written as shown in (40), where $k_{Q_{NR_{21}}}$ is

$$v_1^+(t-\frac{T_m}{4})\times m(t-\frac{T_m}{4})$$
$$=\frac{2}{j\pi}e^{-j\omega\frac{T_m}{4}}\sum_{p=1}^{\infty}\frac{1}{2p-1}(e^{-j\frac{(2p-1)\pi}{2}}e^{-\frac{(2n+1)\pi}{4Q_{NR}}\sqrt{\frac{\omega+(2p-1)\omega_m}{(2n+1)\omega_m}}}e^{j(\omega+(2p-1)\omega_m)t}-e^{j\frac{(2p-1)\pi}{2}}e^{-\frac{(2n+1)\pi}{4Q_{NR}}\sqrt{\frac{|\omega-(2p-1)\omega_m|}{(2n+1)\omega_m}}}e^{j(\omega-(2p-1)\omega_m)t}) \tag{39}$$

$$\begin{aligned} v_2^-(t) &= e^{-j\omega\frac{T_m}{4}}e^{j\omega t}\times\frac{4}{\pi^2}\sum_{p=1}^{\infty}\frac{1}{(2p-1)^2}(e^{-\frac{(2n+1)\pi}{4Q_{NR}}\sqrt{\frac{\omega+(2p-1)\omega_m}{(2n+1)\omega_m}}}+e^{-\frac{(2n+1)\pi}{4Q_{NR}}\sqrt{\frac{|\omega-(2p-1)\omega_m|}{(2n+1)\omega_m}}}) \\ &= e^{-j\omega\frac{T_m}{4}}e^{j\omega t}k_{Q_{NR_{21}}} \\ &= e^{-j\omega\frac{T_m}{4}}v_1^+(t)k_{Q_{NR_{21}}} \end{aligned} \tag{40}$$

the attenuation factor for signals traveling from left to right due to the finite quality factor of the transmission line in the gyrator. For a more accurate expression that includes effects due to the finite Bragg frequency of the line and finite rise and fall times in the modulation signals, the higher inter-modulation terms in (40) should be truncated, and the obtained $(k_{Q_{NR_{21}}})_{truncated}$ should be divided by $k_{filter}$, i.e., $k_{Q_{NR_{21}}} = (k_{Q_{NR_{21}}})_{truncated}/k_{filter}$, so that $k_{Q_{NR_{21}}}$ isolates the effect of the tranmission-line's finite quality factor. Similarly, $k_{Q_{NR_{12}}}$ can be calculated for the signals traveling from right to left.

For the balanced gyrator, $k_{Q_{NR_{12}}}$ and $k_{Q_{NR_{21}}}$ will be different because the signals follow a different path for forward and reverse directions, and the attenuation factor depends on the length traversed. However, for the doubly-balanced gyrator, these factors will be equal, since the phase non-reciprocity is achieved due to sign inversion from the quad-mixer, rather than path length differences. Assuming that the matching of the gyrator has not degraded due to the quality factor of the transmission line, the S-parameter matrix of the gyrator at the center frequency can be written as

$$S=\begin{pmatrix}0 & k_{Q_{NR_{12}}}e^{j\frac{\pi}{2}} \\ k_{Q_{NR_{21}}}e^{-j\frac{\pi}{2}} & 0\end{pmatrix}. \tag{41}$$

From perturbation analysis, the S-parameters of the circulator can be expressed as

$$|S_{21}|\approx\frac{Q_{NR_{effec}}}{Q_{NR_{effec}}+1} \tag{42}$$

$$|S_{32}|\approx\frac{Q_{NR_{effec}}}{Q_{NR_{effec}}+1} \tag{43}$$

where $Q_{NR_{effec}}$ is the effective quality factor associated with the loss of the transmission line in the gyrator, and is given by $2(\frac{1+k_{Q_{NR_{12}}}k_{Q_{NR_{21}}}}{1-k_{Q_{NR_{12}}}k_{Q_{NR_{21}}}}))$.

### *H. Estimating Transmission Losses of the Circulator*

Under the assumption that all the non-idealities are small, the losses due to the individual effects can be added when estimating the final loss of the circulator. Hence, the transmission S-parameters of the circulator can be expressed as

$$|S_{21}|\approx(\frac{Q_1}{Q_1+1})(\frac{2Q_2}{2Q_2+1})(\frac{2Q_3}{2Q_3+1})(\frac{Q_{filter}}{Q_{filter}+1})(\frac{Q_{NR_{effec}}}{Q_{NR_{effec}}+1})(\frac{Q_{sw}}{Q_{sw}+1}) \tag{44}$$

$$|S_{32}|\approx(\frac{2Q_1}{2Q_1+1})(\frac{Q_2}{Q_2+1})(\frac{2Q_3}{2Q_3+1})(\frac{Q_{filter}}{Q_{filter}+1})(\frac{Q_{NR_{effec}}}{Q_{NR_{effec}}+1}). \tag{45}$$

## V. Discrete-Component-Based Case Studies

### *A. Balanced Arbitrary-Phase Non-Reciprocal Element*

A balanced arbitrary-phase non-reciprocal element has been implemented using coaxial cables and commercially-available off-the-shelf switches as shown in Fig. 15. It consists of a two 10m long 50Ω coaxial cables with a time delay of 40.5ns and an insertion loss of 1dB. These cables are sandwiched between two single-pole double-throw (SPDT) switches which are modulated at 6.17MHz. In practice, open-reflective switches with fast switching time are not commercially available. Alternatively, we used short-reflective switches from Minicircuits (model number ZFSW-2-46) with a fast switching time of 2ns, to minimize the insertion loss due to rise/fall time as discussed in Section IV. Using a short-reflective switch does not change the principle of operation of the structure. Signals traveling in the forward direction do not see the short termination. However, signals traveling in the reverse direction experience two reflections from a short termination (Γ=-1) instead of two reflections from an open termination (Γ=1). The two additional negative signs cancel each other, and result in an output signal which is equal to the open-reflective switch case. At 10MHz, the switch has an insertion loss of 0.5dB in the on-state and a reflection coefficient of $0.71\angle 178°$ in the off-state. The modulation signals are generated using arbitrary waveform generators (model: Agilent 33500B) and have a rise/fall time of 8.5ns.

The measured S-parameters between 0-50MHz are shown in Fig. 16. A non-reciprocal phase of 180° has been observed at 6.043MHz, 18.11MHz, 30.2MHz, etc. This shift in the gyration frequency is due to the reflection co-efficient of the off-state switch, which is 178° instead of 180°. Since the reverse traveling wave is reflected twice, a total of -4° phase shift is picked up by the signal. Due to this additional phase shift, the structure behaves as a gyrator at slightly lower frequencies, a similar effect to the observation in [34]. At 10MHz, the transmission losses in forward and reverse directions are -3.8dB and -17dB respectively. The transmission losses in the reverse direction are higher due to the additional loss mechanisms, including 3× loss of the transmission line due to reflections, non-unity reflection co-efficient of the reflective-short switch in its off-state, and destructive interference between the leaked signal to port-1 during off-state (due to imperfect clocking) and the actual signal.

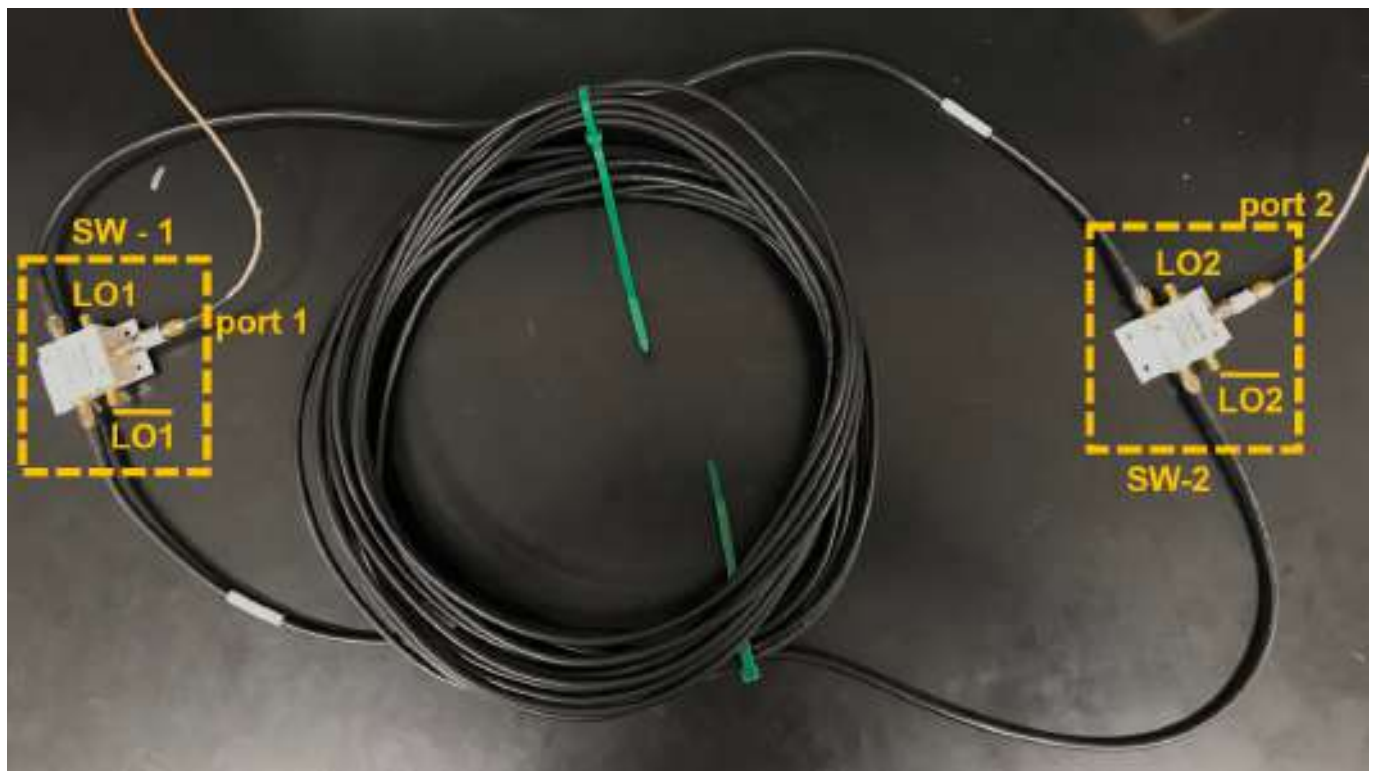

Fig. 15. Photograph of the balanced arbitrary-phase non-reciprocal element assembly.

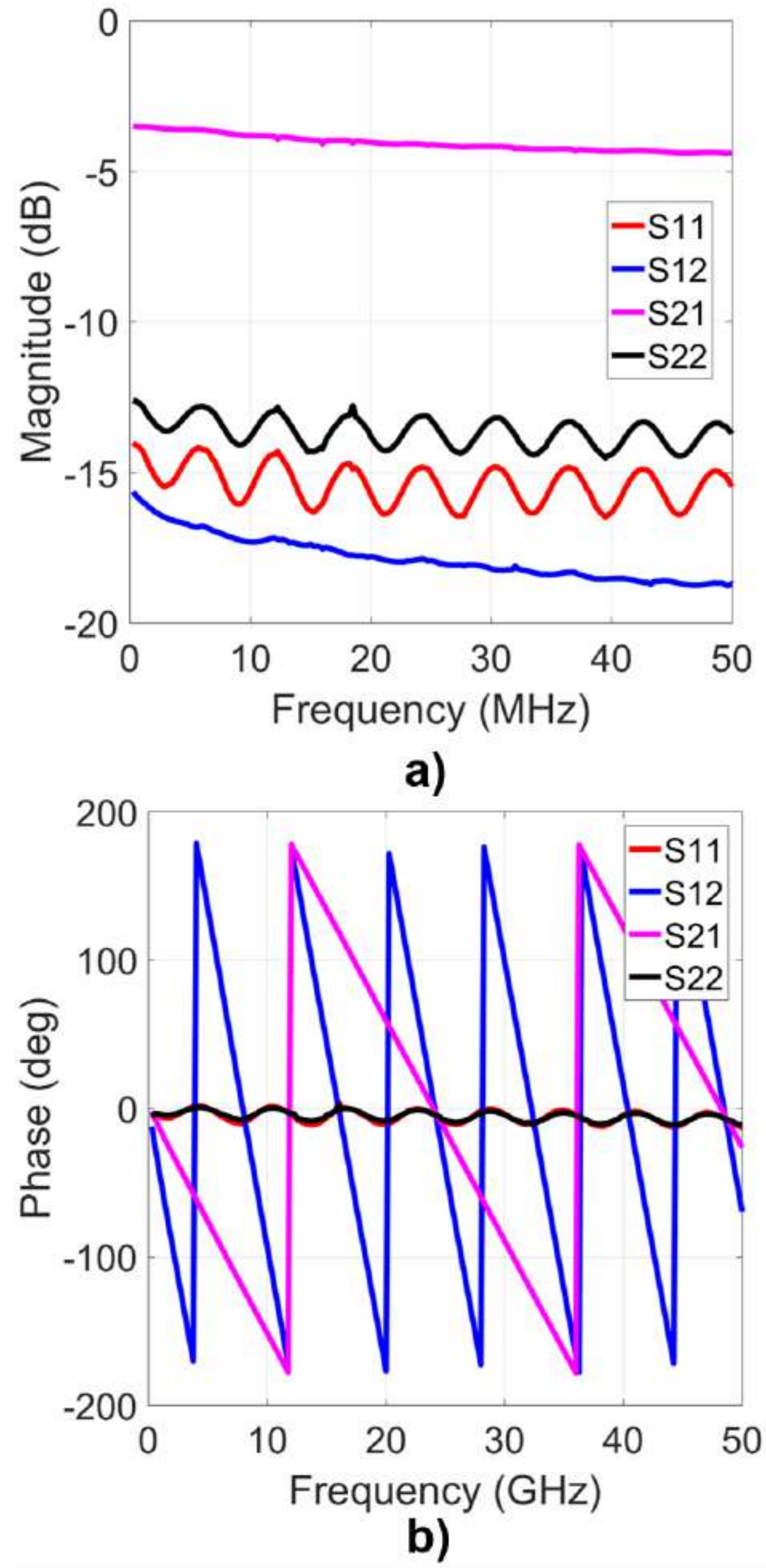

Fig. 16. Measured S-parameters of the balanced arbitrary-phase non-reciprocal element: (a) magnitude, and (b) phase.

### B. Ultra-broadband Dissipative Isolator

An ultra-broadband dissipative isolator has been implemented by the replacing the left-side switch (SW-1) in the balanced configuration setup with a commercially available absorptive switch from Minicircuits (model number: ZFSWA-2-46). The measured S-parameters between 0-50MHz are shown in Fig. 17. At 10MHz, the transmission loss in the forward direction is -3.5dB, and the isolation in the reverse direction is 40dB over the entire frequency range.

## VI. A 25GHz 45nm SOI CMOS Circulator Case Study

### A. Implementation Details

A differential circulator based on the doubly-balanced gyrator was implemented at 25 GHz in 45nm SOI CMOS as depicted in Fig. 18. Aside from achieving superior isolation bandwidth due to the use of the doubly-balanced gyrator, the fully-differential architecture reduces LO feedthrough and enables 3 dB higher power handling at the expense of a doubling in the power consumption. The quarter-wave sections between ports 1 and 2, and 2 and 3, were implemented using differential conductor-backed coplanar waveguides with a Q of 15 at 25 GHz. The gyrator element was placed symmetrically between port 1 and port 3 so that the parasitic capacitances from the mixer switches could be absorbed into the quasi-distributed transmission line in the gyrator and the $\lambda/8$ C-L-C sections on either side. The transmission line in the gyrator is a combination of four $\pi$-type C-L-C sections, with inductor Q of 20, and coplanar waveguides, with a Q of 15, connecting the C-L-C sections. The overall quality factor and Bragg frequency of the transmission line are 17 and 83.9 GHz respectively. The Bragg frequency was improved by a factor of two because only a part of the quarter-period delay required was obtained from the four C-L-C sections, with the rest obtained from the coplanar waveguides used for interconnects. An extra quadrature phase path was added in the gyrator element to counter the loss degradation due to duty-cycle impairments. The reader is directed to [40] for additional details. The gilbert quad-mixers are designed using $2\times16\,\mu m/40\,nm$ floating body transistors, achieving a $R_{sw}$ of $8.66\,\Omega$. The placement of the gyrator symmetrically between port 1 and port 3 distributes the loss due to non-zero switch ON resistance equally between $S_{21}$ ans $S_{32}$, as opposed to (23) and (24). The new transmission losses due to non-zero ON resistance can be expressed as

$$|S_{21}| \approx \frac{2Q_{sw}}{2Q_{sw}+1}, \quad (46)$$

$$|S_{32}| \approx \frac{2Q_{sw}}{2Q_{sw}+1}, \quad (47)$$

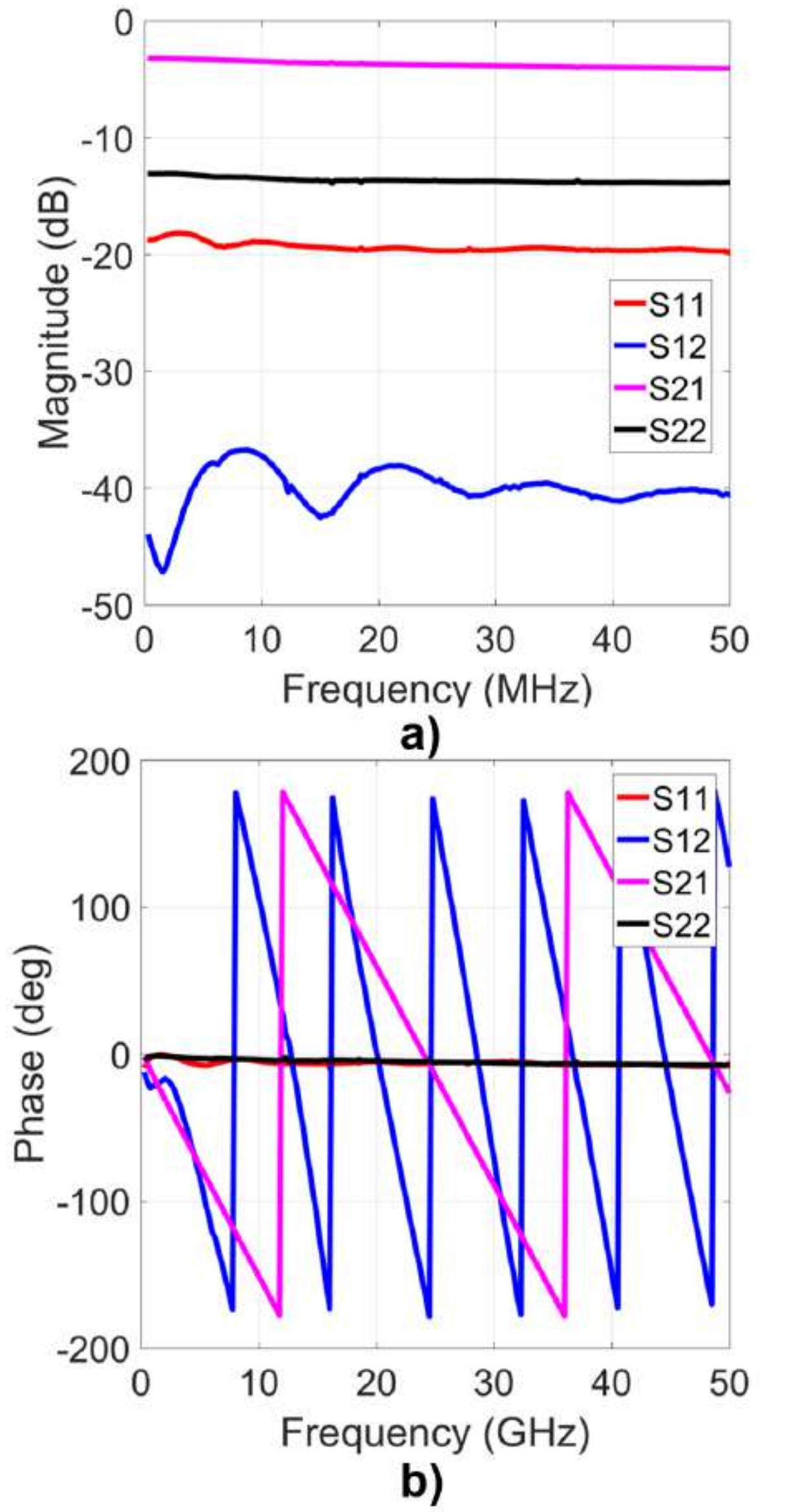

Fig. 17. Measured S-parameters of the ultra-broadband dissipative isolator: (a) magnitude, and (b) phase.

modifying the circulator transmission losses in (44) and (45) to (48) and (49).

Fig. 19 shows the circuit diagram of the 8.33 GHz clock path. A two stage poly-phase filter is used to generate the 8.33 GHz quadrature signals driving the mixer switches. After the poly-phase filter, a three stage self-biased CMOS buffer chain with inductive peaking in the final stage generates the square wave clock signals with a rise/fall time of 7.5 ps for the switches. Independently controlled NMOS varactors at the differential LO inputs provide I/Q calibration of range $\pm 10°$. Interested readers can refer to [40] for more comprehensive implementation details.

### B. Comparison between Theory and Simulations

For this 25 GHz circulator, as mentioned earlier, $Q_1$ and $Q_2$ at 25 GHz are 15 each as the $\lambda/4$ sections were implemented using coplanar waveguides, and $Q_3$ at 25 GHz is 20 as it was implemented using C-L-C sections. $Q_{sw}$ is 5.77 as the single-ended port impedance is 50 Ω and $R_{sw}$ is 8.66 Ω. The quality factor of the delay line in the gyrator, $Q_{NR}$ at 25 GHz, is 17. The Bragg Frequency is 76 GHz and the bandwidth of the modulation signal, $0.35/t_r$, is 46.66 GHz. Hence, the modulation signals are limited to their $5^{th}$ harmonics, as are the inter-modulation terms formed. $k_{filter}$ can therefore be calculated as shown in (50). From (50), $Q_{filter}$ can be calculated to be 14.4. Similarly, $(k_{Q_{NR}})_{truncated}$ can be calculated as shown in (51). From (51), $k_{Q_{NR}}$ can be calculated using $(k_{Q_{NR}})_{truncated}/k_{filter}$. Hence $Q_{NR_{effec}}$ is 7.18. From (44) and (45), $S_{21}$ and $S_{32}$ are -2.66 dB each. The transmission losses, $S_{21}$ and $S_{32}$, at the center frequency from post-layout simulations assuming ideal baluns are -2.8 dB each as shown in Fig. 21. Simulated transmission losses after de-embedding the on-chip baluns are ≈3 dB. Therefore, the simulation results show a close agreement with our analysis.

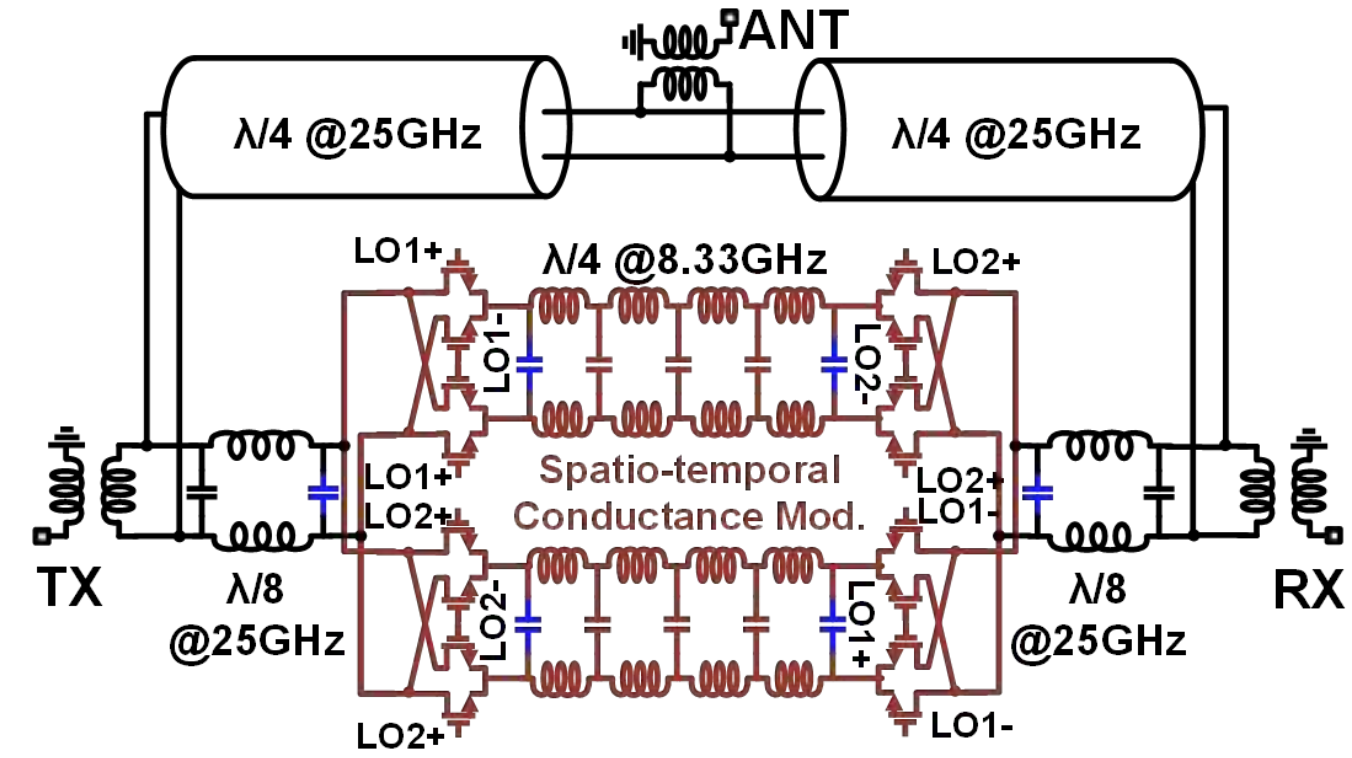

Fig. 18. Circuit diagram of the implemented 25 GHz doubly-balanced circulator

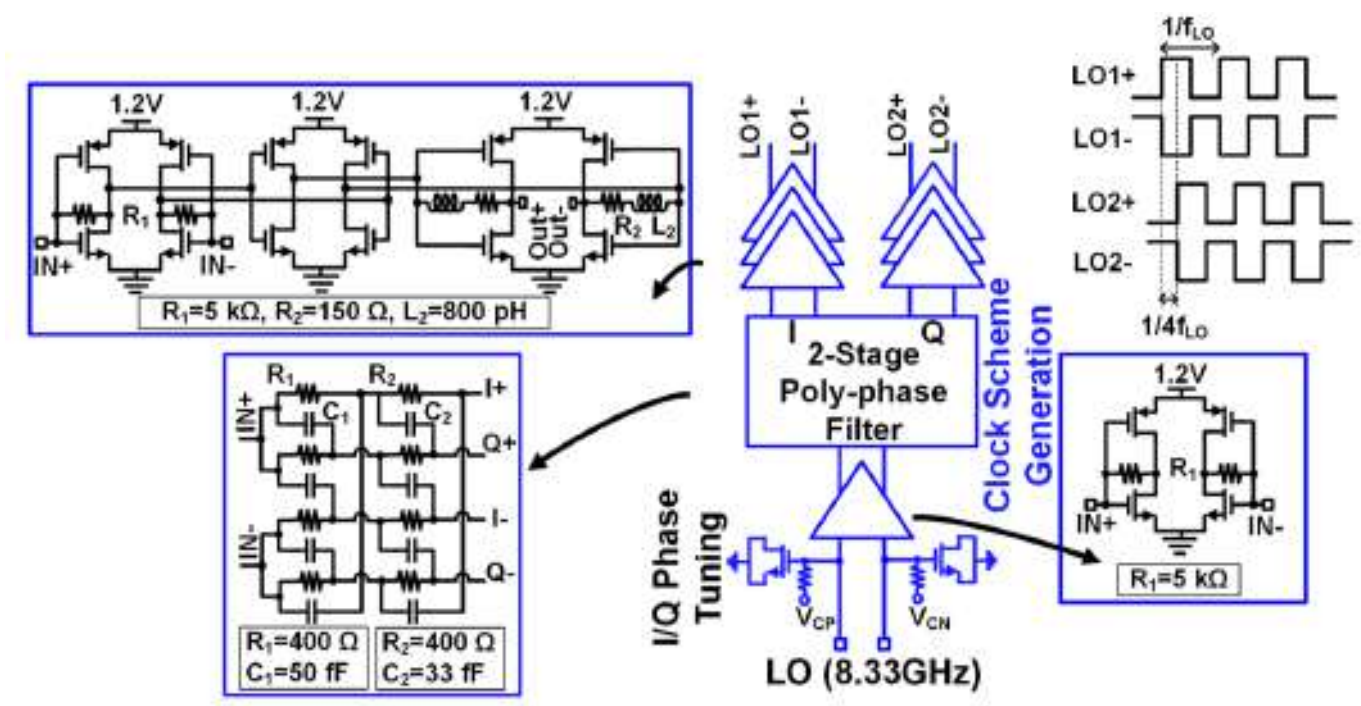

Fig. 19. Circuit diagram of the quadrature modulation signal generation

### C. Measurements

The implemented circulator occupies an area of 1.2 mm×1.8 mm and Fig. 22 shows the die micrograph. The circulator was implemented with on-chip baluns for measurement purposes. The measurements were performed through RF probing in a chip-on-board configuration, and the baluns are de-embedded to obtain the circulator performance. A 180° hybrid (Krytar 4010265) is used to generate the differential (0°/180°) 8.33 GHz signals from a signal generator to drive the clock inputs of the circulator. A two-port Anritsu 37397E Lightning VNA is used to measure the S-parameters

$$|S_{21}| \approx \left(\frac{Q_1}{Q_1+1}\right)\left(\frac{2Q_2}{2Q_2+1}\right)\left(\frac{2Q_3}{2Q_3+1}\right)\left(\frac{Q_{filter}}{Q_{filter}+1}\right)\left(\frac{Q_{NR_{effec}}}{Q_{NR_{effec}}+1}\right)\left(\frac{2Q_{sw}}{2Q_{sw}+1}\right). \tag{48}$$

$$|S_{32}| \approx \left(\frac{2Q_1}{2Q_1+1}\right)\left(\frac{Q_2}{Q_2+1}\right)\left(\frac{2Q_3}{2Q_3+1}\right)\left(\frac{Q_{filter}}{Q_{filter}+1}\right)\left(\frac{Q_{NR_{effec}}}{Q_{NR_{effec}}+1}\right)\left(\frac{2Q_{sw}}{2Q_{sw}+1}\right). \tag{49}$$

$$\begin{aligned} k_{filter}e^{-j\omega t} &= \frac{-4}{\pi^2}[(-e^{j(\omega+\omega_m-\omega_m)t} - e^{j(\omega-\omega_m+\omega_m)t}) + \frac{1}{3^2}(-e^{j(\omega+3\omega_m-3\omega_m)t} - e^{j(\omega-3\omega_m+3\omega_m)t})] \\ &\quad + \frac{1}{5^2}(-e^{j(\omega+5\omega_m-5\omega_m)t} - e^{j(\omega-5\omega_m+5\omega_m)t})] \\ &= 0.933. \end{aligned} \tag{50}$$

$$\begin{aligned} (k_{Q_{NR}})_{truncated}e^{-j\omega t} &= \frac{-4}{\pi^2}[(-e^{-\frac{3\pi}{4Q_{NR}}\sqrt{\frac{\omega+\omega_m}{\omega_m}}}e^{j(\omega+\omega_m-\omega_m)t} - e^{-\frac{3\pi}{4Q_{NR}}\sqrt{\frac{|\omega-\omega_m|}{\omega_m}}}e^{j(\omega-\omega_m+\omega_m)t}) \\ &\quad + \frac{1}{3^2}(-e^{-\frac{3\pi}{4Q_{NR}}\sqrt{\frac{\omega+3\omega_m}{\omega_m}}}e^{j(\omega+3\omega_m-3\omega_m)t} - e^{-\frac{3\pi}{4Q_{NR}}\sqrt{\frac{|\omega-3\omega_m|}{\omega_m}}}e^{j(\omega-3\omega_m+3\omega_m)t})] \\ &\quad + \frac{1}{5^2}(-e^{-\frac{3\pi}{4Q_{NR}}\sqrt{\frac{\omega+5\omega_m}{\omega_m}}}e^{j(\omega+5\omega_m-5\omega_m)t} - e^{-\frac{3\pi}{4Q_{NR}}\sqrt{\frac{|\omega-5\omega_m|}{\omega_m}}}e^{j(\omega-5\omega_m+5\omega_m)t})] \\ &= 0.742. \end{aligned} \tag{51}$$

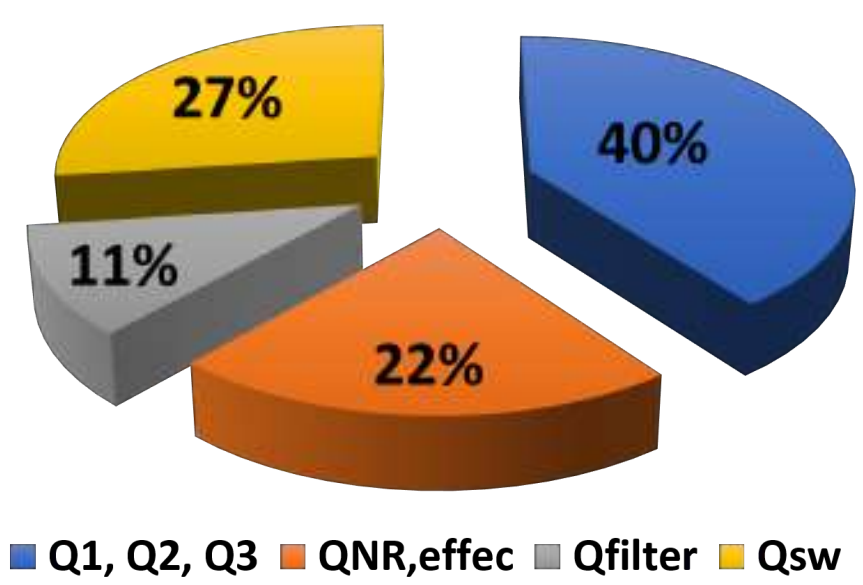

| Non-Ideality | Loss |
|---|---|
| $Q_1$, $Q_2$, $Q_3$ | 1.06 |
| $Q_{NR,effect}$ | 0.58 |
| $Q_{filter}$ | 0.3 |
| $Q_{sw}$ | 0.72 |

**Loss calculated = 2.66 dB**

Fig. 20. Calculated transmission losses, $S_{21}$ and $S_{32}$.

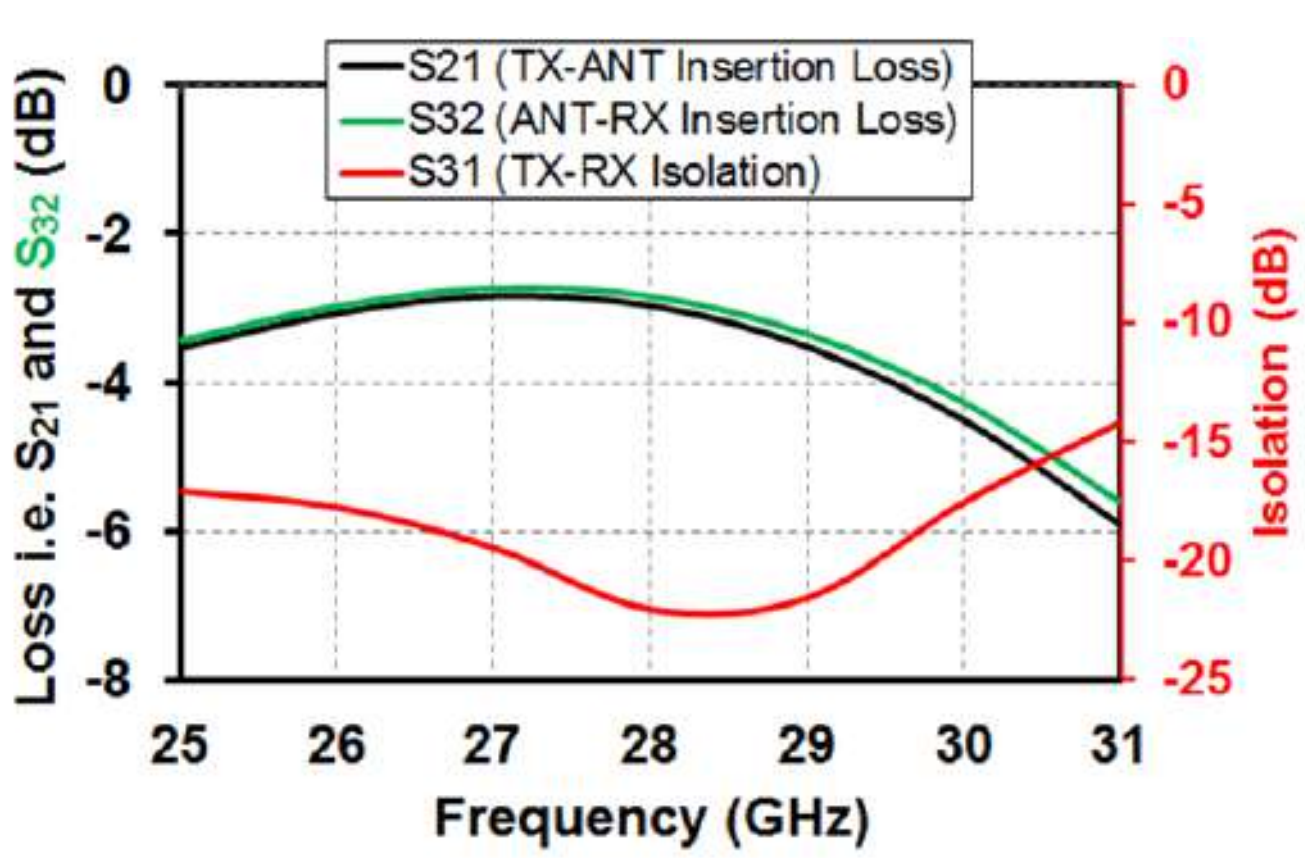

Fig. 21. Post-layout simulations results of transmission losses, $S_{21}$ and $S_{32}$, and isolation $S_{31}$ of the 25 GHz 45 nm SOI CMOS doubly-balanced circulator

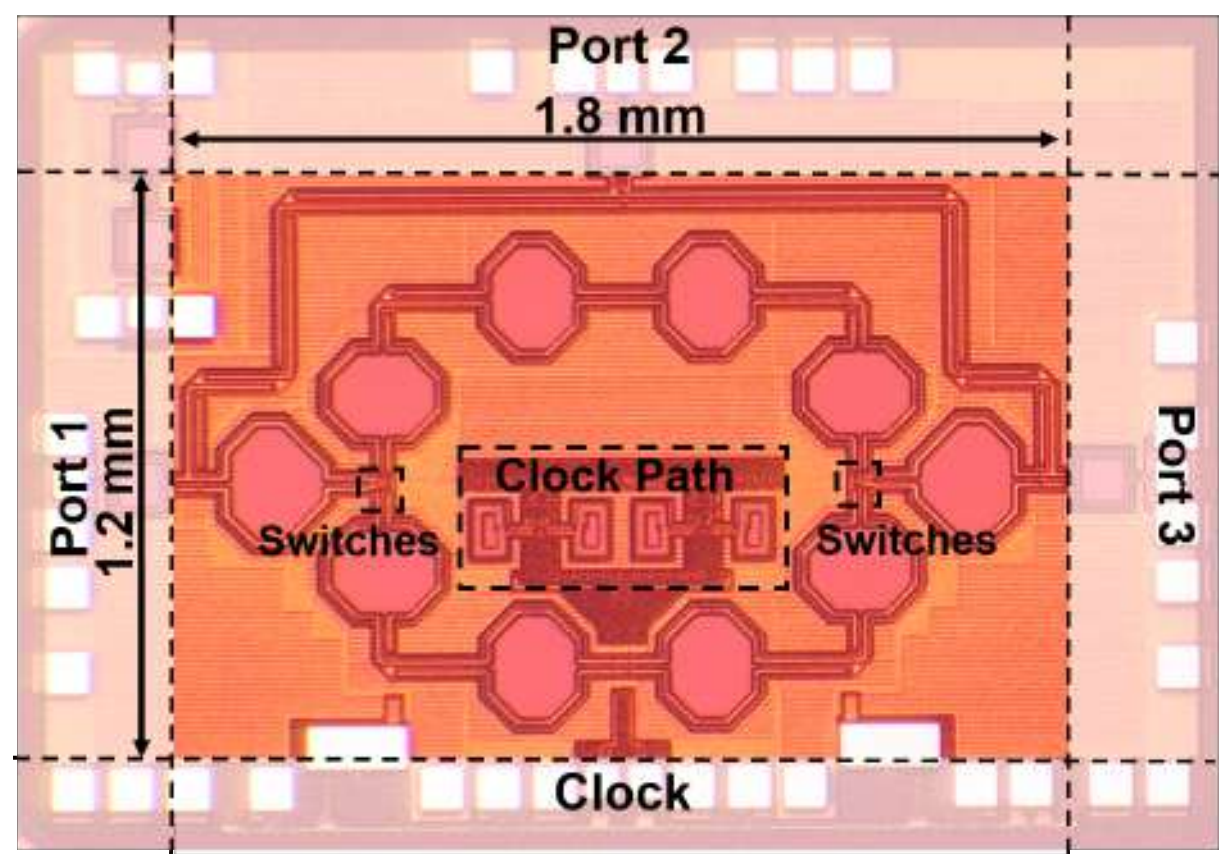

Fig. 22. Die micrograph of the 25 GHz doubly-balanced circulator implemented in the GF 45 nm SOI CMOS process.

by probing two ports at a time, while a millimeter-wave probe terminated with a broadband 50 Ω termination is landed on the third port (Fig. 23).

With the the circulator configured for clockwise circulation with 8.33GHz modulation signals, the measured transmission in clockwise direction, $S_{21}, S_{32}$ and $S_{13}$, shown in Fig. 24, are -3.3 dB, -3.2 dB and -8.7 dB, respectively, without any port impedance tuning. The measured isolations in the reverse direction, $S_{12}, S_{23}$ and $S_{31}$, shown in Fig. 24, are -10.9 dB, -9 dB and -18.9 dB, respectively, without any port impedance tuning. The design of the circulator was optimized to give best performance for $S_{21}$, $S_{32}$ and $S_{31}$ because these (TX-ANT loss, ANT-RX loss and TX-RX isolation) are the most critical parameters in wireless applications.

Importantly, the measured insertion losses are close to the post-layout simulations and the analytical calculations, validating the ability of the analysis presented in this paper to illuminate design trade-offs and provide accurate estimates of performance. The measured isolation of -18.9 dB is limited by the measurement setup. For all circulators, if the ANT port is terminated with an imperfect impedance, the TX-RX isolation will be limited to the return loss at the ANT port. Here, the -18.9 dB isolation is limited by the reflection coefficient of the millimeter-wave probe and its broadband termination. In general, and at millimeter-wave frequencies in particular, an antenna tuner must be tightly integrated with the circulator to obtain best TX-RX isolation.

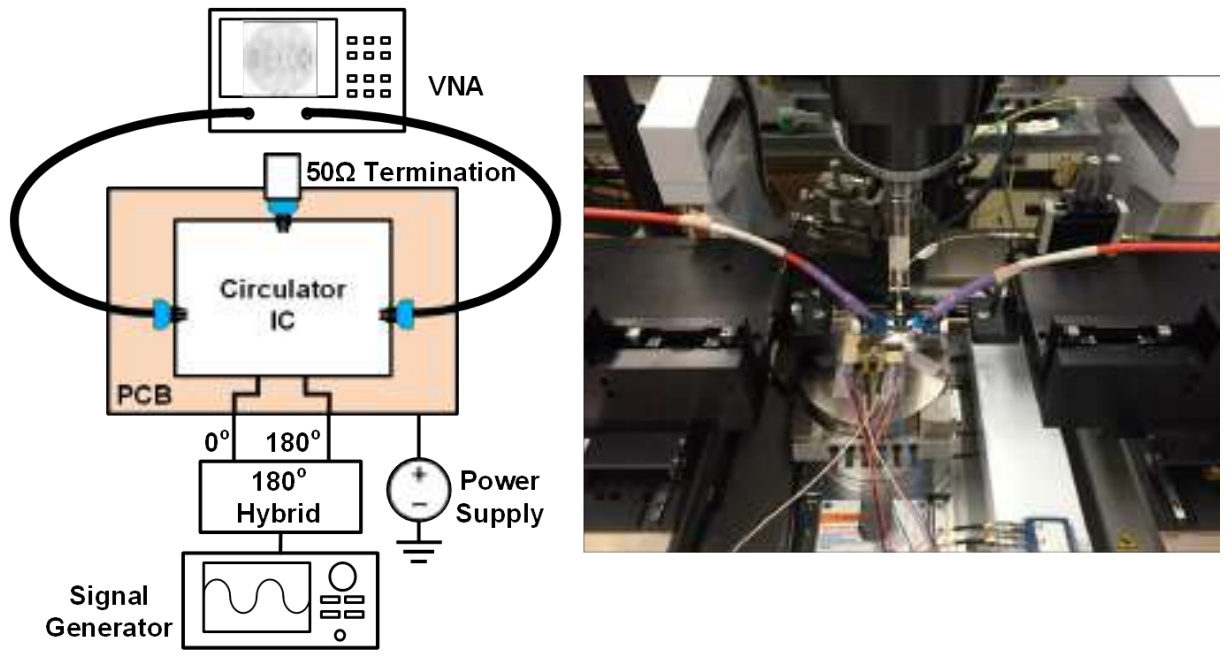

Fig. 23. S-parameter measurement by probing two ports at a time, with a broadband 50 Ω millimeter-wave probe termination on the third port.

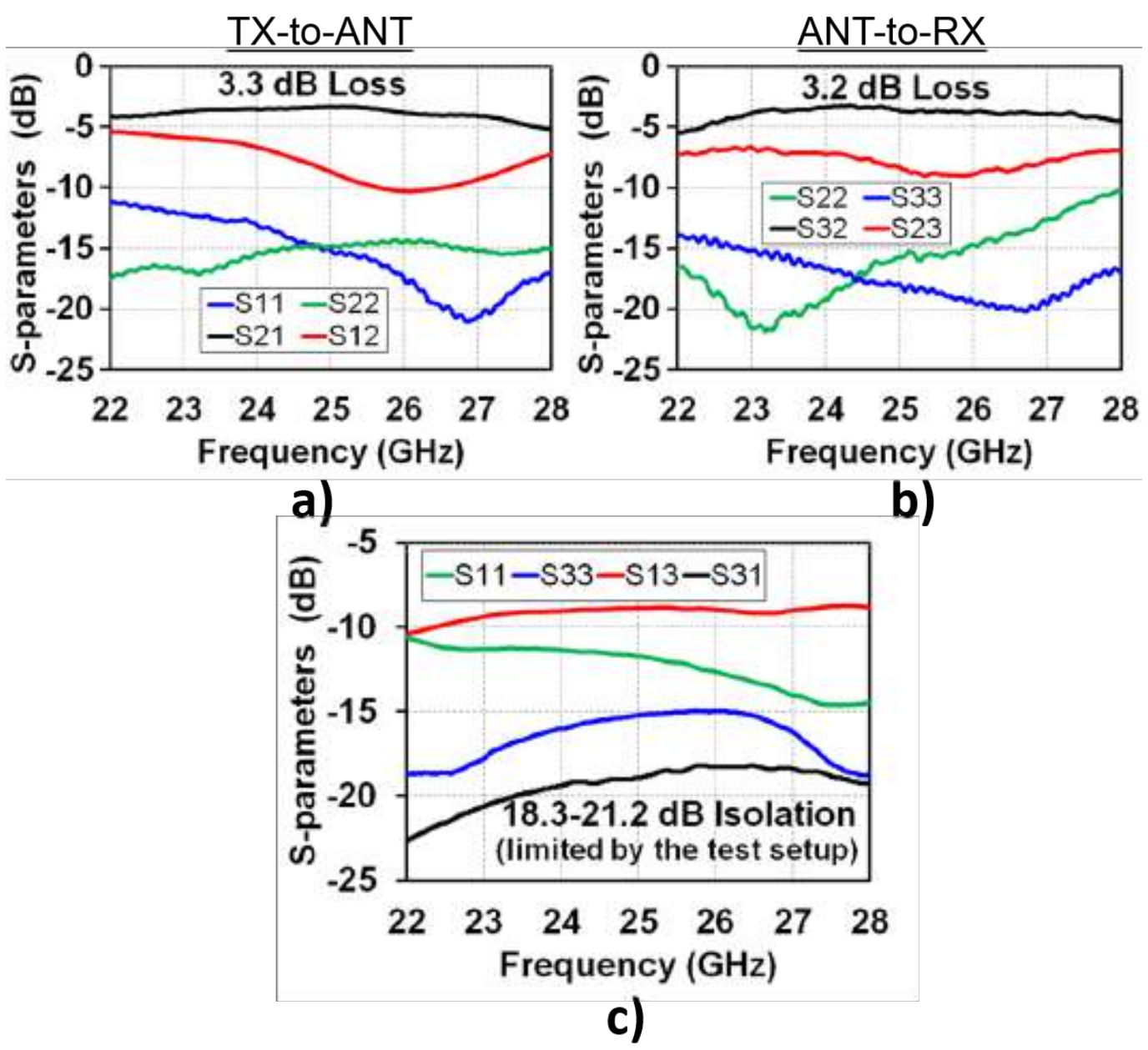

Fig. 24. (a) Measured S-parameters between port 1 and port 2 when port 3 is terminated with a 50 Ω termination. (b) Measured S-parameters between port 2 and port 3 when port 1 is terminated with a 50 Ω termination. (c) Measured S-parameters between port 3 and port 1 when port 2 is terminated with a 50 Ω termination.

## VII. Conclusion

This paper presented various non-reciprocal structures such as a frequency-conversion isolator, a broadband isolator, gyrators, a broadband circulator, and highly-linear single-ended and differential circulators, enabled by using switch-based spatio-temporal conductivity modulation around tranmission lines. The working principle of these structures has been explained using time-domain analysis. In addition, we have also provided a detailed analysis for estimating the transmission losses by factoring in various non-idealities encountered during implementation. We have shown that the theoretically-estimated loss matches closely with our simulation and measurement results. Such an analysis can aid the designer in choosing between different implementation technologies, fabrication processes, and non-reciprocal element topologies, and in optimizing performance by making better trade-offs.

Topics for future research include the discovery of new topologies that further improve insertion loss, isolation bandwidth, and linearity, the incorporation of antenna tuning functionality into the circulator architecture, and investigation and mitigation of the impact of clock phase noise on circulator performance.

## Acknowledgment

This research was sponsored by the DARPA ACT and SPAR programs. The authors would like to thank Dr. Roy Olsson and Dr. Ben Epstein of DARPA for valuable discussions.

## References

[1] D. Bharadia, E. McMilin, and S. Katti, "Full Duplex Radios," in *Proceedings of the ACM SIGCOMM 2013 Conference on SIGCOMM*, ser. SIGCOMM '13, 2013, pp. 375–386.

[2] A. Sabharwal, P. Schniter, D. Guo, D. W. Bliss, S. Rangarajan, and R. Wichman, "In-Band Full-Duplex Wireless: Challenges and Opportunities," *IEEE Journal on Selected Areas in Communications*, vol. 32, no. 9, pp. 1637–1652, Sept 2014.

[3] B. Debaillie, D. J. van den Broek, C. Lavn, B. van Liempd, E. A. M. Klumperink, C. Palacios, J. Craninckx, B. Nauta, and A. Prssinen, "Analog/RF Solutions Enabling Compact Full-Duplex Radios," *IEEE Journal on Selected Areas in Communications*, vol. 32, no. 9, pp. 1662–1673, Sept 2014.

[4] D. Yang, H. Yksel, and A. Molnar, "A Wideband Highly Integrated and Widely Tunable Transceiver for In-Band Full-Duplex Communication," *IEEE Journal of Solid-State Circuits*, vol. 50, no. 5, pp. 1189–1202, May 2015.

[5] D. J. van den Broek, E. A. M. Klumperink, and B. Nauta, "An In-Band Full-Duplex Radio Receiver With a Passive Vector Modulator Downmixer for Self-Interference Cancellation," *IEEE Journal of Solid-State Circuits*, vol. 50, no. 12, pp. 3003–3014, Dec 2015.

[6] D. Korpi, J. Tamminen, M. Turunen, T. Huusari, Y. S. Choi, L. Anttila, S. Talwar, and M. Valkama, "Full-duplex mobile device: pushing the limits," *IEEE Communications Magazine*, vol. 54, no. 9, pp. 80–87, September 2016.

[7] J. Zhou, N. Reiskarimian, J. Diakonikolas, T. Dinc, T. Chen, G. Zussman, and H. Krishnaswamy, "Integrated Full Duplex Radios," *IEEE Communications Magazine*, vol. 55, no. 4, pp. 142–151, April 2017.

[8] N. Reiskarimian, M. B. Dastjerdi, J. Zhou, and H. Krishnaswamy, "Highly-linear integrated magnetic-free circulator-receiver for full-duplex wireless," in *2017 IEEE International Solid-State Circuits Conference (ISSCC)*, Feb 2017, pp. 316–317.

[9] N. Reiskarimian, J. Zhou, and H. Krishnaswamy, "A CMOS Passive LPTV Nonmagnetic Circulator and Its Application in a Full-Duplex Receiver," *IEEE Journal of Solid-State Circuits*, vol. 52, no. 5, pp. 1358–1372, May 2017.

[10] B. D. H. Tellegen, "The Gyrator, A New Electric Network Element," *Philips Res. Rep.*, vol. 3, no. 81, pp. 3003–3014, Dec 1948.

[11] H. J. Carlin, "On the Physical Realizability of Linear Non-Reciprocal Networks," *Proceedings of the IRE*, vol. 43, no. 5, pp. 608–616, May 1955.

[12] R. Y. Barazarte, G. G. Gonzalez, and M. Ehsani, "Generalized Gyrator Theory," *IEEE Transactions on Power Electronics*, vol. 25, no. 7, pp. 1832–1837, July 2010.

[13] P. J. Allen, "The Turnstile Circulator," *IRE Transactions on Microwave Theory and Techniques*, vol. 4, no. 4, pp. 223–227, October 1956.

[14] H. Bosma, "On Stripline Y-Circulation at UHF," *IEEE Transactions on Microwave Theory and Techniques*, vol. 12, no. 1, pp. 61–72, Jan 1964.

[15] S. Tanaka, N. Shimomura, and K. Ohtake, "Active circulators - The realization of circulators using transistors," *Proceedings of the IEEE*, vol. 53, no. 3, pp. 260–267, March 1965.

[16] S. Wang, C. H. Lee, and Y. B. Wu, "Fully Integrated 10-GHz Active Circulator and Quasi-Circulator Using Bridged-T Networks in Standard CMOS," *IEEE Transactions on Very Large Scale Integration (VLSI) Systems*, vol. 24, no. 10, pp. 3184–3192, Oct 2016.

[17] T. Kodera, D. L. Sounas, and C. Caloz, “Artificial faraday rotation using a ring metamaterial structure without static magnetic field,” *Applied Physics Letters*, vol. 99, no. 3, p. 031114, 2011.

[18] T. Kodera, D. Sounas, and C. Caloz, “Magnetless nonreciprocal metamaterial (MNM) technology: Application to microwave components,” *IEEE Transactions on Microwave Theory and Techniques*, vol. 61, no. 3, pp. 1030–1042, March 2013.

[19] G. Carchon and B. Nanwelaers, “Power and noise limitations of active circulators,” *IEEE Transactions on Microwave Theory and Techniques*, vol. 48, no. 2, pp. 316–319, Feb 2000.

[20] B. Peng, S. K. Ozdemir, F. Lei, F. Monifi, M. Gianfreda, G. L. Long, S. Fan, F. Nori, C. M. Bender, and L. Yang, “Paritytime-symmetric whispering-gallery microcavities,” *Nature Physics*, vol. 10, pp. 394–398, April 2014.

[21] L. Fan, J. Wang, L. T. Varghese, H. Shen, B. Niu, Y. Xuan, A. M. Weiner, and M. Qi, “An all-silicon passive optical diode,” *Science*, vol. 335, no. 6067, pp. 447–450, 2012.

[22] K. Gallo, G. Assanto, K. R. Parameswaran, and M. M. Fejer, “All-optical diode in a periodically poled lithium niobate waveguide,” *Applied Physics Letters*, vol. 79, no. 3, pp. 314–316, 2001.

[23] A. M. Mahmoud, A. R. Davoyan, and N. Engheta, “All-passive nonreciprocal metastructure,” *Nature Communications*, vol. 6, no. 8359, July 2015.

[24] S. Qin, Q. Xu, and Y. E. Wang, “Nonreciprocal Components With Distributedly Modulated Capacitors,” *IEEE Transactions on Microwave Theory and Techniques*, vol. 62, no. 10, pp. 2260–2272, Oct 2014.

[25] N. A. Estep, D. L. Sounas, J. Soric, and A. Alu, “Magnetic-free non-reciprocity and isolation based on parametrically modulated coupled-resonator loops,” *Nature Physics*, vol. 10, pp. 923–927, Nov 2014.

[26] N. A. Estep, D. L. Sounas, and A. Al, “Magnetless Microwave Circulators Based on Spatiotemporally Modulated Rings of Coupled Resonators,” *IEEE Transactions on Microwave Theory and Techniques*, vol. 64, no. 2, pp. 502–518, Feb 2016.

[27] D. L. Sounas and A. Alu, “ Non-reciprocal photonics based on time modulation,” *Nature Photonics*, vol. 11, no. 12, Dec 2017.

[28] S. Tyagi, C. Auth, P. Bai, G. Curello, H. Deshpande, S. Gannavaram, O. Golonzka, R. Heussner, R. James, C. Kenyon, S. H. Lee, N. Lindert, M. Liu, R. Nagisetty, S. Natarajan, C. Parker, J. Sebastian, B. Sell, S. Sivakumar, A. S. Amour, and K. Tone, “An advanced low power, high performance, strained channel 65nm technology,” in *IEEE InternationalElectron Devices Meeting, 2005. IEDM Technical Digest.*, Dec 2005, pp. 245–247.

[29] N. Reiskarimian and H. Krishnaswamy, “Magnetic-free non-reciprocity based on staggered commutation,” *Nature Communications*, vol. 7, no. 11217, April 2016.

[30] T. Dinc and H. Krishnaswamy, “A 28GHz magnetic-free non-reciprocal passive CMOS circulator based on spatio-temporal conductance modulation,” in *2017 IEEE International Solid-State Circuits Conference (ISSCC)*, Feb 2017, pp. 294–295.

[31] T. Dinc, M. Tymchenko, A. Nagulu, D. Sounas, A. Alu, and H. Krishnaswamy, “Synchronized Conductivity Modulation to Realize Broadband Lossless Magnetic-Free Non-Reciprocity,” *Nature Communications*, vol. 8, no. 765, Oct 2017.

[32] M. M. Biedka, R. Zhu, Q. M. Xu, and Y. E. Wang, “Ultra-Wide Band Non-reciprocity through Sequentially-Switched Delay Lines,” *Scientific Reports*, vol. 7, no. 40014, Jan 2017.

[33] M. Biedka, Q. Wu, X. Zou, S. Qin, and Y. E. Wang, “Integrated time-varying electromagnetic devices for ultra-wide band nonreciprocity,” in *2018 IEEE Radio and Wireless Symposium (RWS)*. IEEE, 2018, pp. 80–83.

[34] J. Krol and S. Gong, “A non-magnetic gyrator utilizing switched delay lines,” in *2017 47th European Microwave Conference (EuMC)*, Oct 2017, pp. 452–455.

[35] C. Andrews and A. C. Molnar, “A Passive Mixer-First Receiver With Digitally Controlled and Widely Tunable RF Interface,” *IEEE Journal of Solid-State Circuits*, vol. 45, no. 12, pp. 2696–2708, Dec 2010.

[36] A. Ghaffari, E. A. M. Klumperink, M. C. M. Soer, and B. Nauta, “Tunable High-Q N-Path Band-Pass Filters: Modeling and Verification,” *IEEE Journal of Solid-State Circuits*, vol. 46, no. 5, pp. 998–1010, May 2011.

[37] S. Hameed, M. Rachid, B. Daneshrad, and S. Pamarti, “Frequency-Domain Analysis of N-Path Filters Using Conversion Matrices,” *IEEE Transactions on Circuits and Systems II: Express Briefs*, vol. 63, no. 1, pp. 74–78, Jan 2016.

[38] ——, “Frequency-domain analysis of a mixer-first receiver using conversion matrices,” in *2015 IEEE International Symposium on Circuits and Systems (ISCAS)*, May 2015, pp. 541–544.

[39] Q. Zhang, T. Guo, B. A. Khan, T. Kodera, and C. Caloz, “Coupling Matrix Synthesis of Nonreciprocal Lossless Two-Port Networks Using Gyrators and Inverters,” *IEEE Transactions on Microwave Theory and Techniques*, vol. 63, no. 9, pp. 2782–2792, Sept 2015.

[40] T. Dinc, A. Nagulu, and H. Krishnaswamy, “A Millimeter-wave Non-Magnetic Passive SOI CMOS Circulator Based on Spatio-Temporal Conductivity Modulation,” *IEEE Journal of Solid-State Circuits*, vol. 52, no. 4, pp. 3276 – 3292, Dec 2017.